\documentclass[pdflatex,sn-mathphys,referee,a4paper]{sn-jnl}

\usepackage{siunitx}
\usepackage{physics}
\usepackage{graphicx,hyperref,amsmath,bm,url}
\usepackage{cleveref}

\DeclareSIUnit\angstrom{\text {Å}}

\newcommand{\BSO}{B_\mathrm{SO}}
\newcommand{\VLD}{V_\mathrm{LD}}
\newcommand{\VRD}{V_\mathrm{RD}}
\newcommand{\VL}{V_\mathrm{L}}
\newcommand{\VR}{V_\mathrm{R}}
\newcommand{\IL}{I_\mathrm{L}}
\newcommand{\IR}{I_\mathrm{R}}
\newcommand{\Icorr}{I_\mathrm{corr}}
\newcommand{\VPG}{V_\mathrm{PG}}
\newcommand{\muL}{\mu_\mathrm{LD}}
\newcommand{\muR}{\mu_\mathrm{RD}}

\newcommand{\su}{\uparrow} 
\newcommand{\sd}{\downarrow} 
\newcommand{\bpm}{\begin{pmatrix}}
\newcommand{\epm}{\end{pmatrix}}
\newcommand{\nn}{\nonumber \\} 
\newcommand{\tp}{ ^{\intercal} }
\newcommand{\dg}{^{\dagger}}
\newcommand{\hc}{\rm{H.c.}}
\newcommand{\half}{\frac{1}{2}}

\begin{document}

\title[Triplet Cooper pair splitting]{Singlet and triplet Cooper pair splitting in hybrid superconducting nanowires}

\author[1]{\fnm{Guanzhong}~\sur{Wang}}
\equalcont{These authors contributed equally to this work.}

\author*[1]{\fnm{Tom}~\sur{Dvir}}\email{tom.dvir@gmail.com}
\equalcont{These authors contributed equally to this work.}

\author[1]{\fnm{Grzegorz~P.}~\sur{Mazur}}
\equalcont{These authors contributed equally to this work.}

\author[1]{\fnm{Chun-Xiao}~\sur{Liu}}

\author[1]{\fnm{Nick}~\sur{van~Loo}}

\author[1]{\fnm{Sebastiaan~L.~D.}~\sur{ten~Haaf}}

\author[1]{\fnm{Alberto}~\sur{Bordin}}

\author[2]{\fnm{Sasa}~\sur{Gazibegovic}}

\author[2]{\fnm{Ghada}~\sur{Badawy}}

\author[2]{\fnm{Erik~P.~A.~M.}~\sur{Bakkers}}

\author[1]{\fnm{Michael}~\sur{Wimmer}}

\author[1]{\fnm{Leo~P.}~\sur{Kouwenhoven}}

\affil[1]{\orgdiv{QuTech and Kavli Institute of NanoScience}, \orgname{Delft University of Technology}, \postcode{2600 GA} \orgaddress{\city{Delft}, \country{The Netherlands}}}

\affil[2]{\orgdiv{Department of Applied Physics}, \orgname{Eindhoven University of Technology}, \postcode{5600 MB} \orgaddress{\city{Eindhoven}, \country{The Netherlands}}}

\abstract{
In most naturally occurring superconductors, electrons with opposite spins are paired up to form Cooper pairs. 
This includes both conventional $s$-wave superconductors such as aluminum as well as high-$T_\text{c}$, $d$-wave superconductors. 
Materials with intrinsic $p$-wave superconductivity, hosting Cooper pairs made of equal-spin electrons, have not been conclusively identified, nor synthesized, despite promising progress~\cite{ran2019nearly, zhou2021superconductivity, zhou2022isospin}.
Instead, engineered platforms where $s$-wave superconductors are brought into contact with magnetic materials have shown convincing signatures of equal-spin pairing~\cite{robinson2010controlled, khaire2010observation, sprungmann2010evidence}.
Here, we directly measure equal-spin pairing between spin-polarized quantum dots.
This pairing is proximity-induced from an $s$-wave superconductor into a semiconducting nanowire with strong spin-orbit interaction. 
We demonstrate such pairing by showing that breaking a Cooper pair can result in two electrons with equal spin polarization. 
Our results demonstrate controllable detection of singlet and triplet pairing between the quantum dots. 
Achieving such triplet pairing in a sequence of quantum dots will be required for realizing an artificial Kitaev chain~\cite{kitaev2001unpaired, sau2012realizing, leijnse2012parity}.
}

\maketitle

\begin{figure}[htbp]
\centering
\includegraphics[width=\textwidth]{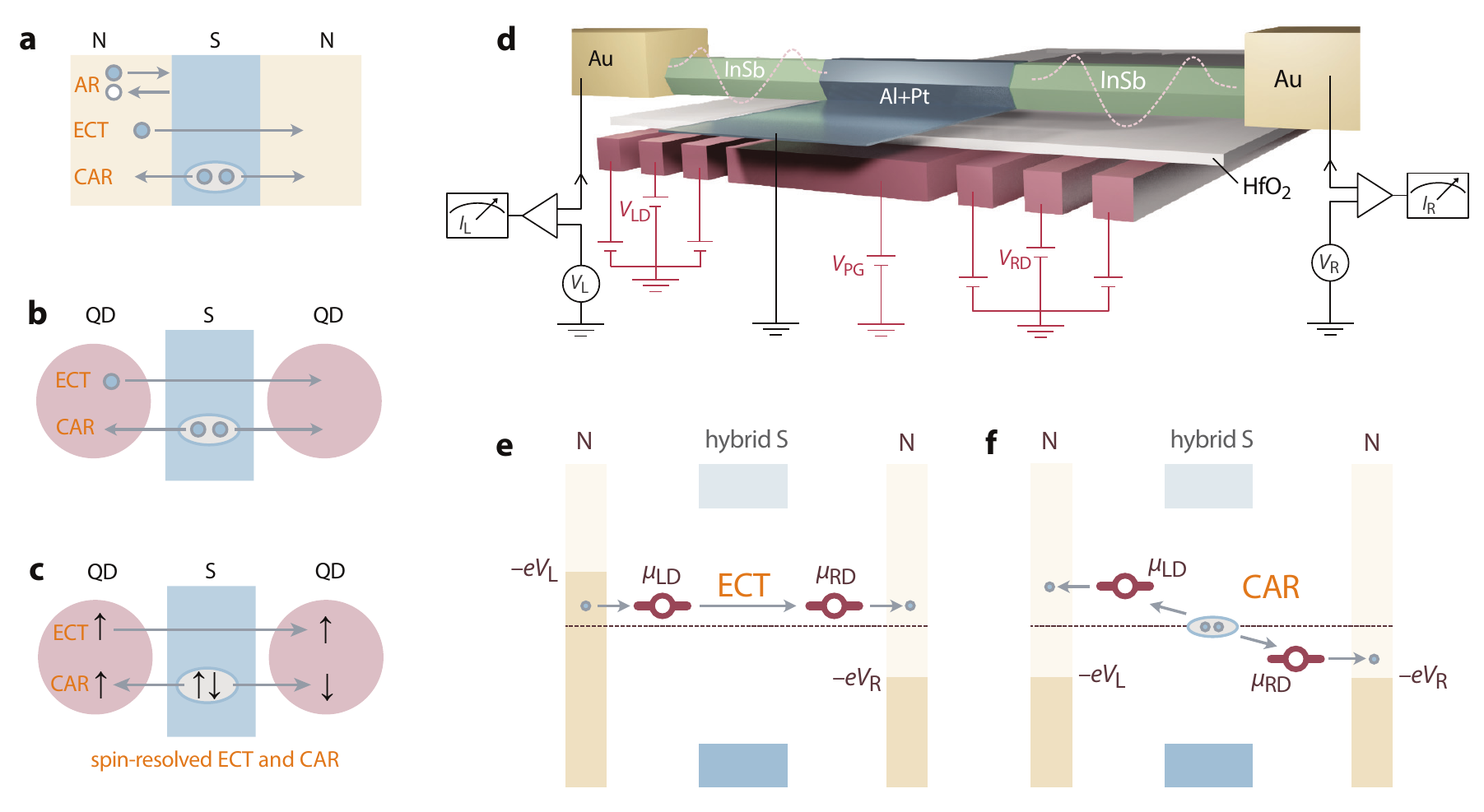}
\caption{\textbf{Transport processes, device and energy diagrams.}
\textbf{a.} Possible electron transport processes in an N-S-N structure: local Andreev reflection (AR), elastic co-tunneling (ECT) and crossed Andreev reflection (CAR).
\textbf{b.} A QD-S-QD structure only allows for ECT and CAR.
\textbf{c.} The QDs become spin-selective in a magnetic field, allowing to distinguish ECT between equal spin states and CAR from a singlet Cooper pair involving opposite spins.
\textbf{d.} Illustration of the N-QD-S-QD-N device and the measurement circuit. Dashed potentials indicate QDs defined in the nanowire by finger gates.
\textbf{e., f.} Energy diagrams for ECT (e) and CAR (f) with detection by varying bias voltages and QD energy alignments. Occupied (unoccupied) states are illustrated by darker (lighter) colors.
}\label{fig:triplet-fig1}
\end{figure}

To probe spin pairing, one can split up a Cooper pair, separate the two electrons and measure their spins. The process to split a Cooper pair is known as crossed Andreev reflection (CAR)~\cite{beckmann2004evidence, russo2005experimental, recher2001andreev}. 
In this process, the two electrons end up in two separated non-superconducting probes (\Cref{fig:triplet-fig1}a), each of these normal (N) probes collecting a single elementary charge, $e$. 
Alternative processes exist such as normal Andreev reflection (AR), with a $2e$ charge exchange between a single normal probe and the superconductor (S), and elastic co-tunneling (ECT), with $1e$ charge from one normal probe crossing the superconductor and ending up in the other normal probe. 
AR does not allow to measure the separate spins and thus this process needs to be suppressed. 
Following the approach of previous Cooper pair splitting studies~\cite{hofstetter2009cooper, herrmann2010carbon, das2012high-efficiency, schindele2012near-unity, tan2015cooper, borzenets2016high}, we realize this by using quantum dots (QDs) with large charging energies that only allow for $1e$ transitions. 
This suppresses $2e$-AR to $\sim 5\%$ of the total current in each junction (see \Cref{fig:triplet-ED_linecuts}).
The remaining CAR and ECT processes are sketched in \Cref{fig:triplet-fig1}b. 
In ECT, $1e$ is subtracted from one QD and added to the other, whereas in CAR, an equal-sign $1e$ charge is either added or subtracted simultaneously to each QD. 
We will use this difference to distinguish ECT from CAR. 
Besides charge detection, QDs can be configured to be spin-selective in a magnetic field~\cite{recher2000quantum, hanson2004semiconductor}. 
\Cref{fig:triplet-fig1}c illustrates that ECT involves equal spin states in both QDs, whereas CAR from a singlet Cooper pair requires opposite spin states. 
Interestingly, these rules of spin combinations can be relaxed in the presence of inhomogeneous magnetic fields or spin-orbit interaction, both of which allow the possibility for triplet pairing~\cite{Gorkov2001, annunziata2012proximity, bergeret2014spin, linder2015superconducting, banerjee2018controlling, jeon2020tunable, cai2021evidence, ahmad2021coexistence, phan2022detecting}.  For instance, spin-orbit (SOC) coupling can rotate an opposite-spin configuration into an equal-spin pair. 
In this report, we first demonstrate charge measurements, as illustrated in \Cref{fig:triplet-fig1}b, followed by spin-selective detection of ECT and CAR, which sets us up to detect CAR with equal-spins when spin precessions are induced by SOC.

\section{Charge filtering}

\begin{figure}[htbp]
\centering
\includegraphics[width=0.95\textwidth]{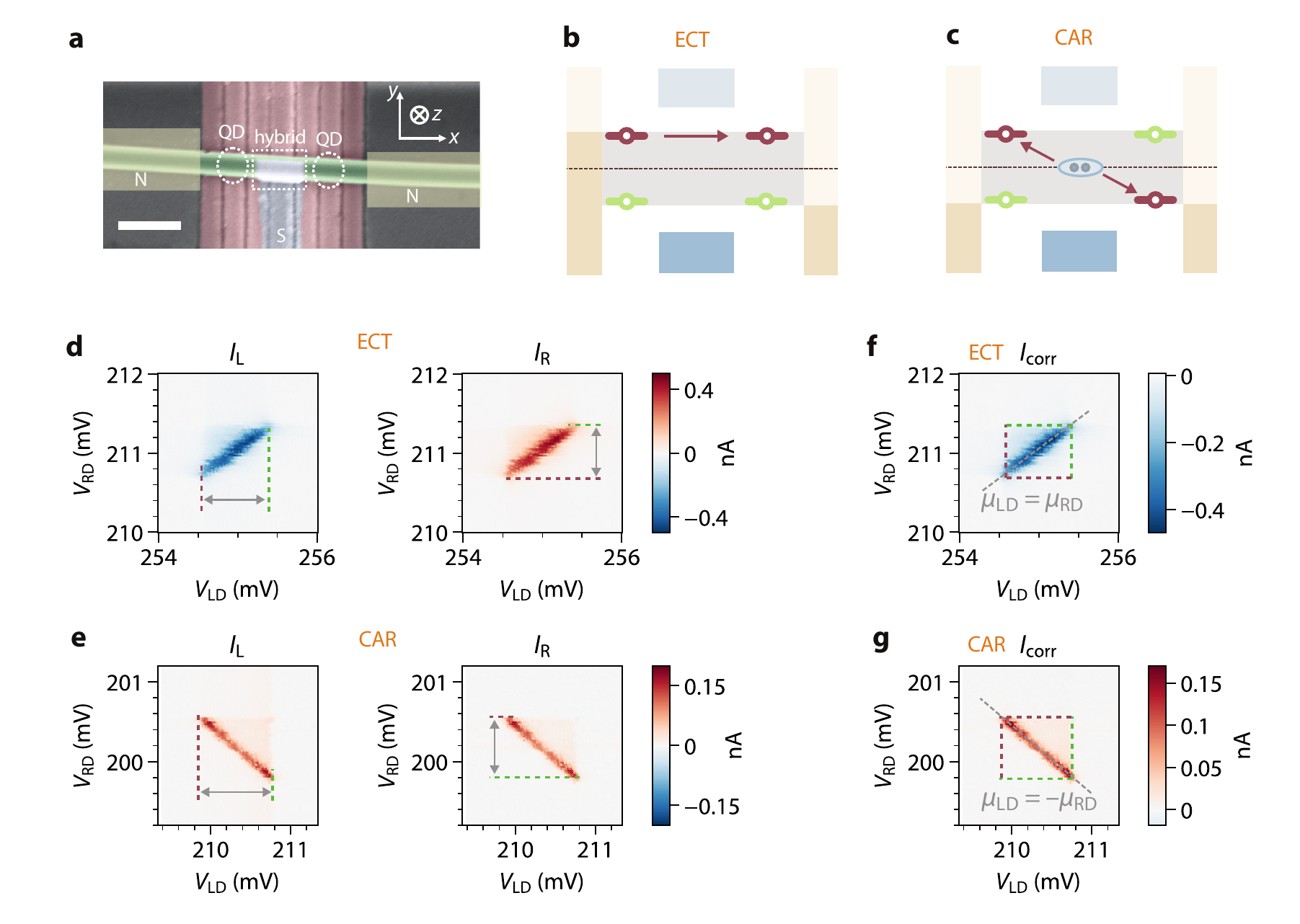}
\caption{\textbf{CAR and ECT.}
\textbf{a.} False-colored SEM image of Device~A prior to the fabrication of N leads, using the same color representation as in \Cref{fig:triplet-fig1}d. Translucent rectangles indicate locations of N leads. Dotted lines indicate QDs and the hybrid segment in the middle. Scale bar is \SI{300}{nm}. Inset: our coordinate system. The nanowire lies at a $3^\circ$ angle to the $x$ axis.
\textbf{b., c.} Energy diagrams for ECT (b) and CAR (c) measurements.  The grey areas bound by bias voltages indicate the transport window. The QD levels represent two possible scenarios of energy alignment at the boundaries of the transport window (brown and green).
\textbf{d.} Measured $\IL$ and $\IR$ for the bias configuration illustrated in panel~b, which selects for ECT. Dashed lines mark the transport window boundaries using the same colors as QD illustrations in panel~b. 
Data was taken with $B_x=\SI{0.1}{T}$ and $\VPG = \SI{0.18}{V}$.
\textbf{e.} Measured $\IL$ and $\IR$ for the bias configuration illustrated in panel~c, which selects for CAR. 
\textbf{f.} The correlated current, $\Icorr$, through the two QDs calculated from data in panel~d. The dashed box marks the transport window, and the diagonal dashed line indicates where the QD levels are aligned. 
\textbf{g}. Idem for data from panel~e. Here, the diagonal dashed line indicates where the QD levels are anti-aligned.
}\label{fig:triplet-fig2}
\end{figure}

The device and the measurement setup are illustrated in \Cref{fig:triplet-fig1}d. A short segment of an InSb nanowire is proximitized by a thin Al shell, which is kept grounded throughout the experiment. Two QDs are formed on both sides of the hybrid segment. The electrochemical potentials in the two QDs, $\muL$ and $\muR$, are controlled by voltages on the respective gates, $\VLD$ and $\VRD$. Crucially, the level spacing between QD orbitals exceeds \SI{1}{meV}, such that near each charge degeneracy the QD can be considered as a single orbital level.  Two normal leads (Au) are attached to both QDs. Both leads are independently voltage biased ($\VL, \VR$), and the currents through the leads are measured separately ($\IL, \IR$).

The energy diagram in \Cref{fig:triplet-fig1}e illustrates that ECT requires alignment of the QD levels ($\muL = \muR$), both positioned within the transport window defined by the bias voltages $\VL$ and $\VR$. 
We restrict the bias settings to $\VL = -\VR$ for ECT unless mentioned otherwise. 
In \Cref{fig:triplet-fig1}e, the transport window is thus defined by $-e \VL > \muL = \muR > -e \VR$. 
To study co-tunneling processes which only occupy a higher-energy intermediary state virtually, the QD excitations and bias voltages are kept within the induced superconducting gap, i.e., lower in energy than any state in the hybrid (see \Cref{fig:triplet-ED_ABS_B}). 
We define current to be positive when flowing from N into S for both sides, implying that ECT yields opposite currents, $\IL = -\IR$. 
On the other hand, CAR requires anti-symmetric alignment between the two QD levels, $\muL = -\muR$~\cite{tan2015cooper}, to satisfy overall energy conservation, as shown in \Cref{fig:triplet-fig1}f. 
We restrict bias settings to $\VL = \VR$ for CAR unless specified. 
Thus, the transport window in \Cref{fig:triplet-fig1}f is now defined by $-e \VL = -e \VR  < \muL = -\muR < e \VL = e \VR $, allowing tunneling from the QDs into empty states in the nearby leads. In our definition the CAR-induced currents are equal: $\IL = \IR$. 
The boundaries of the transport window are further illustrated in \Cref{fig:triplet-fig2}b,~c.

A scanning electron microscope image of the main device, A, is shown in \Cref{fig:triplet-fig2}a. 
In \Cref{fig:triplet-fig2}d we show $\IL$ and $\IR$ as function of the two QD voltages for fixed $\VL = -\VR = \SI{100}{\micro V}$. 
The two currents are close to the expected $\IL = -\IR$ (see also \Cref{fig:triplet-ED_linecuts}) and are strong along a straight line with a positive slope. 
Using the lever arm of QD gates extracted in \Cref{fig:triplet-ED_dotchar}, we find this line to be $\muL = \muR$. 
In panel~e, we set $\VL = \VR = \SI{150}{\micro V}$ and similarly observe $\IL \approx \IR$ along a straight line with a negative slope where $\muL = -\muR$. 
Several features in these data allow us to attribute the origin of these sub-gap currents to CAR and ECT instead of competing transport processes.  
The non-local origin of the measured currents, expressed by the (anti-)symmetric energy requirement on both QDs and current correlation, rules out local Andreev reflection. 
The bias and QD energies being kept lower than any sub-gap bound state excludes resonant tunneling into and out of them. 
The only mechanisms known to us that can explain these observations are CAR and ECT~\cite{kleine2009contact}.

In \Cref{fig:triplet-ED_linecuts} we extract from this measurement Cooper pair splitting visibilities of 91\% and 98\% for the left and right QDs, respectively.
Their product of 90\%, for the first time realized, exceeds the minimum value of 71\% required for a Bell test~\cite{schindele2012near-unity}. 
The high efficiency of Cooper pair splitting reported in this work compared to previous reports relies on having a hard superconducting gap in the proximitized segment and on having multiple gates for each QD, allowing control of the chemical potential of QDs independently from QD-lead couplings. 
Both requirements are enabled by recent advancements in the fabrication technique~\cite{heedt2021shadow}. 
The dashed lines in \Cref{fig:triplet-fig2}d,e indicate the boundaries of the transport window, as illustrated with corresponding colors surrounding the grey area in \Cref{fig:triplet-fig2}b,c. 
For convenience, we introduce the correlated current $\Icorr \equiv \mathrm{sgn}(\IL \IR)\sqrt{\vert \IL \IR\vert }$, plotted in \Cref{fig:triplet-fig2}f,g for the corresponding ECT and CAR measurements. 
This product is finite only when currents through both junctions are nonzero, allowing us to focus on features produced by ECT or CAR (see \Cref{fig:triplet-ED_raw-1,fig:triplet-ED_raw-2}). 
Its sign directly reflects the dominant process: ECT being negative and CAR positive.

\section{Spin blockade at zero magnetic field}

\begin{figure}[h!]
\centering
\includegraphics[width=0.95\textwidth]{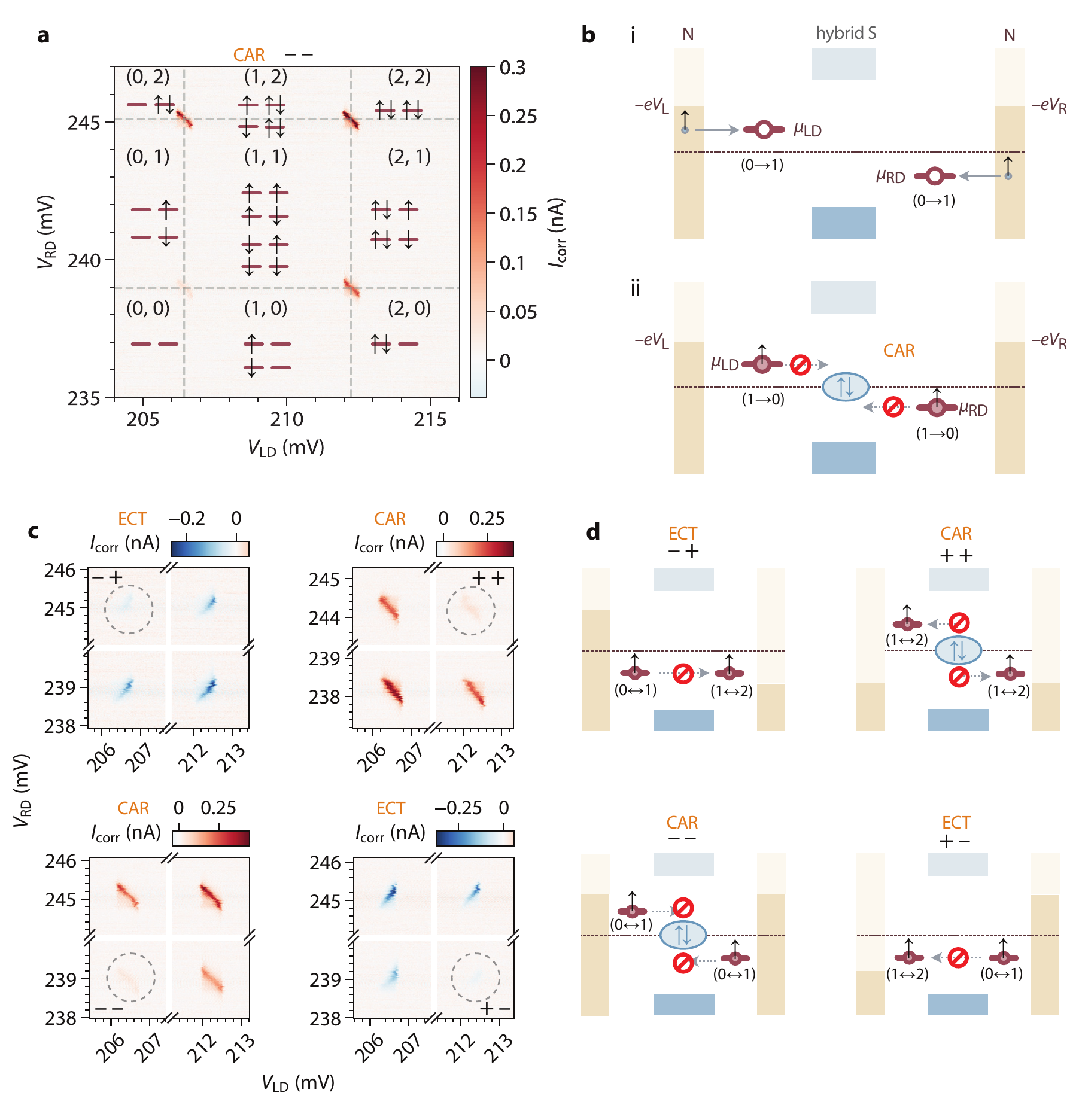}
\caption{\textbf{Spin blockade of CAR and ECT at zero magnetic field.}
\textbf{a.} Charge stability diagram of the two QDs coupled via the hybrid section. $\Icorr$ is measured with \SI{-100}{\micro V} biases on both N leads (denoted $--$), i.e., under CAR conditions. The shown region in gate space covers four charge degeneracy points.
\textbf{b.} Illustration of CAR-mediated spin-blockade. Small arrows indicate the spin polarization of the electrons participating in the transport.
\textbf{c.} $\Icorr$ under all four combinations of bias polarities ($-+, ++, +-, --$ in clockwise order from top left). Axes are interrupted in the Coulomb blockaded range to allow zooming in on charge transitions. Gray dashed circles indicated conditions for either ECT- or CAR-mediated spin-blockade.
\textbf{d.} Illustration of spin-blockade conditions with bias configurations corresponding to the four panels in c. The QD occupations through a complete cycle are indicated.
}\label{fig:triplet-fig3}
\end{figure}

Spin-degenerate orbital levels can each be occupied with two electrons with opposite spins. 
\Cref{fig:triplet-fig3}a shows the charge stability diagram measured with negative biases on both N leads. 
We label the charge occupations relative to the lower-left corner, with some unknown but even number of electrons in each QD. 
Increasing the gate voltages $\VLD$ and $\VRD$ increases the occupation of left and right QD levels one by one from (0,0) to (2,2). In between charge transitions, the occupation is fixed with possible spin configurations as indicated in \Cref{fig:triplet-fig3}a. 
At charge degeneracies, $\Icorr$ is generally non-zero. 
However, the correlated current is very weak at the $(0,0) \leftrightarrow (1,1)$ transition compared to the other three. 
This can be understood as a CAR-mediated spin-blockade, illustrated in panel~b. 
At the $(0,0) \leftrightarrow (1,1)$ transition, each QD can receive an electron with any random spin orientation from the leads. 
Opposite spins can recombine into a Cooper pair. 
However, whenever the QDs are both occupied with the same spin, CAR is suppressed and thereby blocks the transport cycle. 
Note that SOC in InSb is known to not lift this blockade~\cite{danon2009pauli, nadj-perge2010disentangling}. 
\Cref{fig:triplet-fig3}c also shows a similar ECT-mediated spin-blockade when applying anti-symmetric biases to the N leads. 
This effect is intimately related to the well-known Pauli spin-blockade in double QDs~\cite{Ono.2002,hanson2007spins,Hofmann.2017} and shows that spins are well defined and relax slowly compared to the transport cycle time (a few nanoseconds for currents on the order of \SI{100}{pA}).
\Cref{fig:triplet-fig3}c shows CAR and ECT for all four bias-polarity combinations. 
In each of them, one out of the four joint charge degeneracy points exhibits suppressed current. 
The spin configurations that lead to blockade are sketched in panel~d. 
To sum up the general principle, ECT cannot occur if an electron of a certain spin needs to tunnel into an orbital already occupied with the same spin. 
On the other hand, CAR cannot proceed if Cooper pairs must be split into or combined from an equal-spin occupation of the two dots. 
Similar to double QDs, we believe that the residual current under blockade conditions is due to hyperfine interaction~\cite{nadj-perge2010disentangling}.

\section{Spin filtering}

\begin{figure}[h!]
\centering
\includegraphics[width=0.9\textwidth]{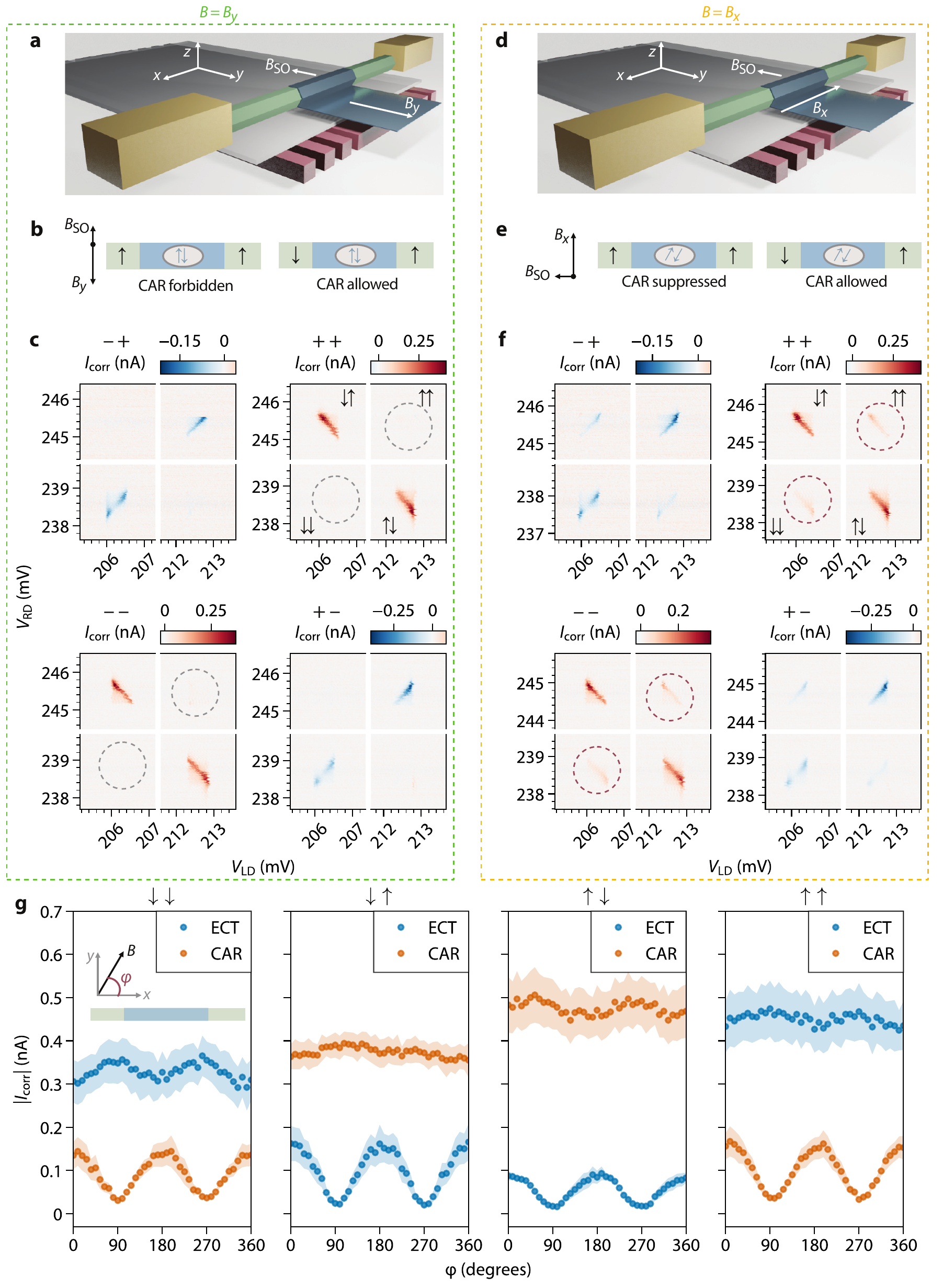}
\caption{\textbf{Spin-resolved CAR and ECT.}
\textbf{a.} Illustration of the externally applied $B$ direction relative to the nanowire axis for panels~b and c. In all illustrations, notations $\uparrow$ and $\downarrow$ are defined along the applied $B$ direction. 
\textbf{b.} Illustration of allowed and forbidden CAR when $B \parallel \BSO$. 
\textbf{c.} $\Icorr$ under the four bias and four spin combinations in the two QDs for $B \parallel \BSO$ and $B = B_y = \SI{100}{mT}$. Vanishing CAR currents due to QD spin filtering are indicated with dashed circles. 
\textbf{d.} Illustration of the $B$ direction for panels~e and f. 
\textbf{e.} Illustration of spin filtering when $B \perp \BSO$. 
\textbf{f.} Similar to panel~c, but for $B=B_x \perp \BSO$. CAR currents that result from equal-spin Cooper pair splitting are circled in red. 
\textbf{g.} $\Icorr$ for four QD spin combinations versus angle of the in-plane magnet field. Each data point is the mean of all four bias configurations and the error bars show the $1\sigma$ spread. Inset: the direction of $B$ relative to the nanowire.
}\label{fig:triplet-fig4}
\end{figure}

At finite magnetic field $B$, the four charge degeneracies in \Cref{fig:triplet-fig3}a can become bipolar spin filters~\cite{recher2000quantum, hanson2004semiconductor}. 
This requires the Zeeman energy in the QDs to exceed the bias voltage, electron temperature, and hyperfine interaction, yet remain smaller than the level spacing of the QDs. 
Under these conditions, we use $\uparrow/\downarrow$ (along the applied $B$ direction) to denote the two spin-split QD eigenstates. 
Only $\downarrow$ electrons are transported across a QD at the $0 \leftrightarrow 1$ transition and only $\uparrow$ electrons at $1 \leftrightarrow 2$. 
\Cref{fig:triplet-fig4}b illustrates the consequence of spin filtering for CAR processes, namely a complete suppression for parallel spins. 
The opposite is expected for ECT with only spin-conserved tunneling being allowed. 
We first apply $B=B_y = \SI{100}{mT}$, in the plane of the substrate and perpendicular to the nanowire.
The four panels in \Cref{fig:triplet-fig4}c present $\Icorr$ measured at four bias polarity combinations, selecting either CAR or ECT conditions. 
The upper right panel also shows the lowest-energy spin combinations.
$\Icorr$ vanishes for $\uparrow \uparrow$ and $\downarrow \downarrow$ with CAR biases $--$ and $++$, and for $\uparrow \downarrow$ and $\downarrow \uparrow$ with ECT biases $+-$ and $-+$. 
The observation of spin conservation suggests spin is a good quantum number.
Thus, any spin-orbit field in the InSb nanowire, $\BSO$ (including both possible Rashba and Dresselhaus SOC), must be parallel, or nearly parallel, to $B_y$.
In this case, CAR provides a coupling mechanism only for an opposite-spin configuration in the two QDs.
We note that the exact $\BSO$ direction as measured by suppression of equal-spin CAR or opposite-spin ECT depends on gate settings and the device used (e.g., \Cref{fig:triplet-ED_devB_rot}).
We have measured directions within $20^\circ$ of being perpendicular to the nanowire axis but its angle with the substrate plane can range from 0 to $60^\circ$.
This observation is consistent with the expectation of $\BSO$ being perpendicular to the nanowire axis for both Rashba and Dresselhaus SOC~\cite{nadj-perge2010spinorbit, Hofmann.2017, Wang.2018}.

To quantify the observation that CAR is anti-correlated with the spin alignment of the QDs, we perform a spin correlation analysis~\cite{Braunecker2013,Busz2017} similar to that in Ref~\cite{bordoloi2022spin}, which analogously reports reduced CAR amplitudes when QD spins are parallel compared to anti-parallel.
The results, presented in \Cref{fig:triplet-Spin_corr}, show two QD spins are anti-correlated by a factor of $-0.86$ for CAR signals when pairing is singlet, the highest reported to date.

When we apply $B \perp \BSO$, in a classical analogy, the spin-orbit interaction leads to spin precession about the $\BSO$ axis in the hybrid section while the QDs remain approximately polarized along $B$~\cite{Stano.2005,Hofmann.2017} (see Supplementary Material for detailed discussions). 
Now, an injected $\uparrow$ electron can acquire a finite $\downarrow$ component and combine with another $\uparrow$ electron into a Cooper pair, as illustrated in \Cref{fig:triplet-fig4}e. 
Similarly, spin precession generates a non-zero probability to couple opposite spins via ECT. 
These expectations are indeed confirmed in \Cref{fig:triplet-fig4}f. 
We again use biases to select ECT or CAR for the four spin-polarized charge degeneracy points. 
Remarkably, faint but finite CAR signals appear in $\uparrow \uparrow$ and $\downarrow \downarrow$ spin combinations (highlighted by dashed circles), as well as for $\uparrow \downarrow$ and $\downarrow \uparrow$ in ECT. 
The observed CAR coupling for $\uparrow \uparrow$ and $\downarrow \downarrow$ is interpreted as a measure of the equal-spin coupling between the QDs. 
In \Cref{fig:triplet-ED_B_sweep}, we show that these observations do not qualitatively depend on the magnitude of $\vert B \vert$ as long as spin polarization is complete (above $\sim \SI{50}{mT}$).

To further investigate the field-angle dependence, we measure CAR and ECT while rotating $\vert B \vert = \SI{100}{mT}$ in the plane of the substrate, see \Cref{fig:triplet-fig4}g. 
For this measurement, we apply a $\pm \SI{100}{\micro V}$ bias voltage only on one side of the device while keeping the other bias zero. 
This allows us to measure both CAR and ECT without changing the applied biases, as can be understood from the same basic principles outlined in \Cref{fig:triplet-fig1} (see \Cref{fig:triplet-ED_devB_rot} for details of this measurement scheme and the affiliated data repository for plots of the raw data). 
We take the maximum value of each $\Icorr$ scan at a particular bias and spin combination as the CAR magnitude and the absolute value of the minimum for ECT. 
Along the two directions parallel to $\BSO$, $\varphi \approx 90^\circ$ and $270^\circ$, equal-spin CAR and opposite-spin ECT are forbidden. 
(The finite extracted amplitudes in this dataset are our noise floor, although small amounts of equal-spin CAR and opposite-spin ECT even when $B \parallel \BSO$ can also be observed in other datasets such as \Cref{fig:triplet-ED_devB_rot}.) 
When $B \perp \BSO$ (i.e., $\varphi \approx 0^\circ$ and $180^\circ$) the anomalous signals are the largest, as expected for effects caused by spin-orbit interaction~\cite{Hofmann.2017, Wang.2018}. 
The signals corresponding to favored spin combinations (e.g., $\uparrow \downarrow$ CAR) do not always exhibit as clear oscillations for reasons we do not yet understand.

\section{Discussion}
The oscillating CAR signals in \Cref{fig:triplet-fig4}g for the two equal-spin configurations (leftmost and rightmost panels) are the central results of this work. 
The presence of the anomalous equal-spin CAR signal shows how non-collinear $B$ and $\BSO$ leads to unconventional spin pairing between QDs. 
Below we discuss possible microscopic scenarios giving rise to SOC-induced spin precession.
InSb nanowires have both Rashba- and Dresselhaus-type SOC. 
Both terms are linear in the momentum along the nanowire axis and their addition gives an effective spin-orbit term in a direction generally perpendicular to the nanowire axis~\cite{Wang.2018}.
Such SOC also exists in our InSb-based QDs and can lead to nominally $\uparrow$ QD eigenstates having a small $\downarrow$ component~\cite{nadj-perge2012spectroscopy}.  
In \Cref{fig:triplet-ED_QD_SOI} and Supplementary Material, we quantify this effect and argue that the opposite-spin admixture is too small to explain the measured amplitude of the ECT and CAR anisotropy. 

The superconducting pairing in the hybrid segment itself is predicted to hold a triplet component due to SOC as well~\cite{alicea2010majorana}. 
The shape and amplitude of our observed oscillations allow comparison with a theory adopting this assumption~\cite{liu2022tunable}, resulting in an estimated spin-orbit strength in the hybrid section between 0.11 and \SI{0.18}{eV \cdot \angstrom} for Device~A and 0.05 to \SI{0.07}{eV \cdot \angstrom} for Device~B (see \Cref{fig:triplet-ED_theory}).
This estimation agrees with reported values in the literature~\cite{de2018electric, bommer2019spin}. 
While the existence of triplet pairing component in the hybrid is thus consistent with our results, it is not the only possible explanation. 
During the tunneling process between the QDs, the electrons traverse through a bare InSb segment, whose SOC could also result in spin precession~\cite{Hofmann.2017, Wang.2018}.
Both scenarios, however, support an interpretation of spin-triplet superconducting coupling between the QDs necessary for construction of a Kitaev chain~\cite{sau2012realizing}.

Finally, we remark that the role of the middle Al-InSb hybrid segment of our devices in electron transport has not been discussed in this work. 
\Cref{fig:triplet-ED_ABS_B} shows that this segment hosts discrete Andreev bound states due to strong confinement in all three dimensions and these states are tunnel-coupled to both N leads. 
The parallel theoretical work modelling this experiment~\cite{liu2022tunable} shows that these states are expected to strongly influence CAR and ECT processes upon variation of the gate voltage underneath the hybrid segment. 
The experimental observations of the gate tunability of CAR and ECT is presented in Ref.~\cite{bordin2022controlled}.

\section{Conclusion}
In conclusion, we have measured CAR and ECT in an N-QD-S-QD-N device with and without spin filtering. 
For well-defined, specific settings consistent with our expectations, we observe Cooper pair splitting for equal spin states in the QD probes. 
These observations are consistent with the presence of a triplet component in the superconducting pairing in the proximitized nanowires, which is one of the building blocks for a topological superconducting phase~\cite{lutchyn2010majorana, oreg2010helical}.
More generally, our results show that the combination of superconductivity and SOC can generate triplet CAR between spin-polarized QDs, paving the road to an artificial Kitaev chain~\cite{kitaev2001unpaired, leijnse2012parity, sau2012realizing}. 
The realization of a Kitaev chain further requires increasing the coupling strength between QDs to allow the formation of a hybridized, extended state. 
This is confirmed in a parallel work where the QDs are driven to the strong coupling regime~\cite{dvir2022realization}.

\section{Methods}

\subsection{Device characterization and setup}

The main device, A, and the measurement setup are illustrated in \Cref{fig:triplet-fig1}d. An InSb nanowire is in ohmic contact with two Cr/Au normal leads. The center is covered with a \SI{200}{nm}-wide thin Al film. Device~A has a \SI{2}{\angstrom}, sub-monolayer Pt grown on top, which increases the magnetic-field compatibility~\cite{mazur2022spin}. 
Device~B presented in \Cref{fig:triplet-ED_devB_rot,fig:triplet-ED_devB_char} has no Pt top layer and has a \SI{350}{nm}-wide middle hybrid segment. 
The Al superconducting lead both proximitizes and grounds the hybrid nanowire segment. The two N leads are independently voltage biased ($\VL, \VR$) and the currents are measured separately ($\IL, \IR$).
Measurements are done at \SI{20}{mK} in a dilution refrigerator using a standard dc transport setup (see below). 
An \SI{18}{nm} layer of HfO$_2$ dielectric separates the nanowire from seven Ti/Pd bottom gates. 
Three gates each in the left and right N-S junctions are used to define QDs. 
The electrochemical potentials in the two QDs, $\muL$ and $\muR$, are controlled by voltages on the respective middle gates, $\VLD$ and $\VRD$. 
Voltage on the central plunger gate, $\VPG$, remains zero for Device~A and \SI{0.4}{V} for Device~B unless mentioned otherwise. 
An SEM image of Device~A is shown in \Cref{fig:triplet-fig2}a.
Characterization of the left and right QDs in Device~A (\Cref{fig:triplet-ED_dotchar}) shows charging energies of \SI{2.1}{meV} and \SI{2.75}{meV} respectively, much larger than the superconducting gap $\Delta = \SI{270}{\micro eV}$ in Al. 
The QDs exhibit irregular Coulomb peak spacings that are typical of the few-electron regime. 
Transport in the N-QD-S junctions is blocked at energies below $\Delta$, confirming strong suppression of AR. 
We note that screening due to the presence of multiple metallic gates and a superconducting film in between diminishes cross-coupling between $\VLD$ and $\VRD$.

\subsection{Device fabrication}

Our hybrid-nanowire devices are fabricated on pre-patterned substrates, following the shadow-wall lithography technique described in Refs~\cite{heedt2021shadow, borsoi2021single} and specific details in the supplementary information of Ref~\cite{mazur2022spin}. 
InSb (111) nanowires are deposited onto the substrates using an optical nanomanipulator setup. 
For Device A, \SI{8}{nm} of Al was grown at a mix of $15^\circ$ and $45^\circ$ angles with respect to the substrate. 
Subsequently, it was coated with \SI{2}{\angstrom} of Pt deposited at $30^\circ$ angle before capping it with \SI{20}{nm} evaporated AlO$x$. 
For Device~B, the same recipe was used with the exception that no Pt coating was deposited. 
Details of the surface treatment of the nanowires, the growth conditions of the superconductor, the thickness calibration of the Pt coating and the ex-situ fabrication of the ohmic contacts can be found in Ref~\cite{mazur2022spin}.

\subsection{Transport measurements}

Devices A and B are cooled down in dilution refrigerators with base temperature $\sim \SI{20}{mK}$, equipped with 3D vector magnets and measured using standard voltage-biased dc circuits illustrated in \Cref{fig:triplet-fig1}. 
No lock-in technique is used except \Cref{fig:triplet-ED_ABS_B}. 
Current amplifier offsets are calibrated using known zero-conductance features when the device is pinched off or in deep Coulomb blockade. 
Total series resistance in each fridge line is \SI{1.85}{k\ohm} for Device~A and \SI{2.9}{k\ohm} for Device~B. 
Total resistance of the voltage source and current meter is $<\SI{0.1}{k\ohm}$ for Device~A and \SI{102}{k\ohm} for Device~B, i.e., much smaller than the device resistance.

We measured six samples fabricated using similar recipes. 
Most devices in these samples suffered from shorts between finger gates or between gates and contacts, possibly due to electrostatic discharge. 
Devices~A and B are the only two we have measured with three functional ohmic contacts, at least six functional finger gates and stable gate dielectric, allowing us to define QDs on both sides. 
Both devices show qualitatively the same behavior.

\subsection{Device tune-up}

The tuning of our device, in particular the QDs, is done as follows. 
First, we form a single barrier between N and S by applying a low voltage on the gate closest to S on each side. 
We then perform local and non-local tunnel spectroscopy of the hybrid segment and locate a $\VPG$ range in which a hard gap is observed at low energies and extended Andreev bound states are observed at high energy (see \Cref{fig:triplet-ED_ABS_B}). 
Having located a desired value of $\VPG$, we form a second barrier in each junction by applying a lower voltage on the gates closest to the N leads. 
The confined region between the two barriers thus becomes a QD. 
We characterize the QDs by measuring its current above the superconducting gap, applying $\vert \VL \vert, \vert \VR \vert > \Delta/e$ as a function of $\VLD,\VRD$ and applied magnetic field (see \Cref{fig:triplet-ED_dotchar}c,d). 
We look for a pair of resonances that correspond to the filling of a single non-degenerate orbital. 
This is indicated by two resonances separated by only the charging energy at zero field and their linear Zeeman splitting when $B>0$. 
We finally measure CAR and ECT between the two QDs (as discussed in \Cref{fig:triplet-fig2}). 
We optimize the measurement by controlling the gates separating the QDs from S to balance low local Andreev current (lowering gate voltage) with high signal-to-noise ratio (raising gate voltage). 
Having reached a reasonable balance, we characterize again the QDs (\Cref{fig:triplet-ED_dotchar}). 

\subsection{Analysis of the structure of the obtained CAR and ECT patterns}
Fitting the data in \Cref{fig:triplet-fig2}d and \Cref{fig:triplet-fig2}e to a theory model~\cite{liu2022tunable} (see supplementary information) yields QD-QD coupling strengths on the order of electron temperature.
Such weak tunnel coupling does not alter the QD eigenstates significantly and allows us to operate QDs as good charge and spin filters.
We further notice that finite ECT and CAR currents can be observed when both QDs are within the transport window but not on the diagonal lines dictating energy conservation. 
Since they appear only on one side of the (anti-)diagonal line corresponding to down-hill energy relaxation, these currents result from inelastic processes involving spontaneous emission and are thus non-coherent.
We note that the data shown in \Cref{fig:triplet-fig2}d and \Cref{fig:triplet-fig2}e are taken at different gate settings than the rest of the paper and are selected because of high data resolution and Cooper pair splitting efficiency.
The (anti-) diagonal resonance line and the strongly (anti-)correlated currents are generic to all QD orbitals that we have investigated. 

\subsection{Role of the Pt layer}
Another source for SOC in our Device~A could come from the Pt sub-atomic top layer, although we have not found evidence for this in previous studies~\cite{mazur2022spin}. 
Note that the spin-orbit scattering in Pt is isotropic and cannot give rise to the angular magnetic field dependence. 
Nevertheless, we have reproduced all the CAR and ECT observations in a second device (Device~B) where the Pt layer was not included (\Cref{fig:triplet-ED_devB_rot,fig:triplet-ED_devB_char}).

\pagebreak

\section{Extended data}

Raw data presented in this work and the processing/plotting codes are available at \url{https://doi.org/10.5281/zenodo.5774827}.

\renewcommand\thefigure{ED\arabic{figure}}
\setcounter{figure}{0}


\begin{figure}[h!]
\centering
\includegraphics[width=0.9\textwidth]{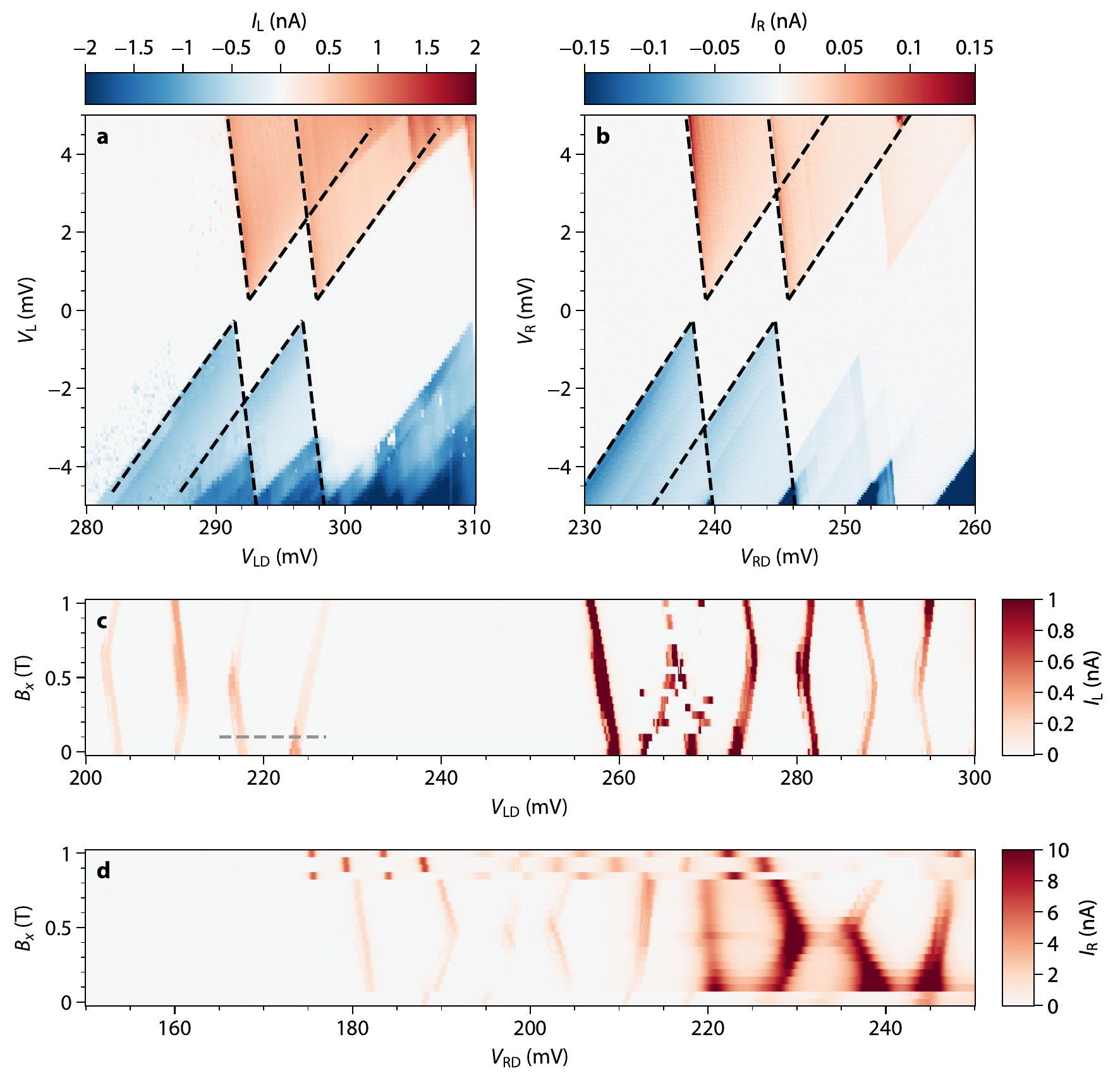}
\caption{\textbf{QD characterization in Device~A}.
\textbf{a.} Coulomb blockade diamonds of the left QD. Superimposed dashed lines represent a model with charging energy \SI{2.1}{meV}, $\Delta=\SI{250}{\micro V}$ and lever arm 0.4.
$\VLD$ shown here is different from other measurements of this resonance due to a drift in one tunnel barrier gate during the process of the experiment.
\textbf{b.} Coulomb blockade diamonds of the right QD. Superimposed dashed lines represent a model with charging energy \SI{2.75}{meV}, $\Delta=\SI{250}{\micro V}$ and lever arm 0.435. In both QDs, no sub-gap current is visible, indicating QDs are weakly coupled to S and retain their charge states.
\textbf{c.} Current through the left QD at $\VL=\SI{500}{\micro V}$ measured against gate voltage and magnetic field along the nanowire, $B_x$.
Spin-degenerate orbitals Zeeman-split in opposite directions while $0< B_x < \SI{0.5}{T}$ and cross around \SI{0.5}{T} when Zeeman energy becomes greater than the level spacing $\sim \SI{1.2}{meV}$ (see \Cref{fig:triplet-ED_QD_SOI} for $g$-factor extraction). The orbital used in \Cref{fig:triplet-fig3,fig:triplet-fig4} is the pair of resonances marked by grey dashed lines at $B=\SI{100}{mT}$.
\textbf{d.} Current through the right QD at $\VR=\SI{500}{\micro V}$. The orbital used in \Cref{fig:triplet-fig3,fig:triplet-fig4} is outside the measured range in this plot immediately to the right. All QD resonances we investigated behave similarly including those in \Cref{fig:triplet-fig2}, which are selected because of high data resolution and Cooper pair splitting efficiency.
}\label{fig:triplet-ED_dotchar}
\end{figure}


\begin{figure}[h!]
\centering
\includegraphics[width=\textwidth]{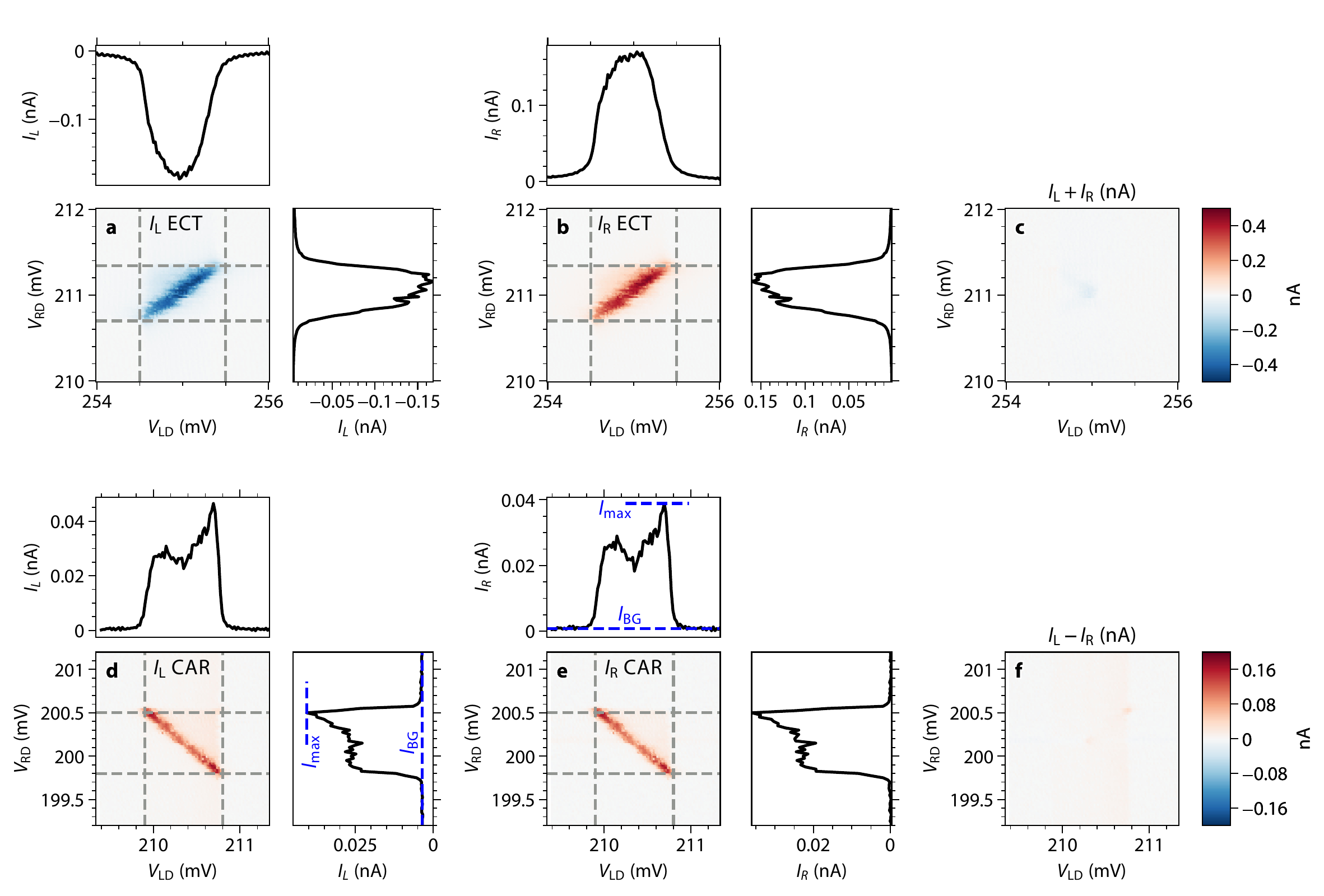}
\caption{\textbf{More analysis on data presented in \Cref{fig:triplet-fig2}, including Cooper pair splitting efficiency extraction at $B=0$.}
\textbf{a., b.} ECT $\IL, \IR$ and averaged currents. Top panel shows signals between the horizontal grey lines averaged over $\VRD$. Right panel shows signals between the vertical grey lines averaged over $\VLD$. Almost no background current is visible unless both dots participate in transport.
\textbf{c.} $\IL+\IR$ of the ECT measurement is almost 0, verifying $\vert\IL\vert = \vert \IR \vert$ in most of the phase space except when both QDs are at zero energy and charge selection no longer plays a role.
\textbf{d., e.} CAR $\IL, \IR$ and averaged currents, similar to panels~a,~b. Using $\eta_\mathrm{L}\equiv (1-I_\mathrm{L,BG}/I_\mathrm{L,max})$ where the background $I_\mathrm{L,BG}$ is taken as the average current when $\VRD$ is off-resonance and $\VLD$ is on-resonance in the right panel of d, we obtain a Cooper pair splitting visibility of 91.3\% for the left junction. Similarly, the right junction has splitting visibility $\eta_\mathrm{R}=98\%$. This gives combined visibility $\eta_\mathrm{L} \eta_\mathrm{R}=89.5\%$.
\textbf{f.} $\IL-\IR$ of the CAR feature is almost 0, verifying $\vert\IL\vert = \vert \IR \vert$ except a small amount of local Andreev current in the left QD, manifesting as a vertical feature independent of $\VRD$ near $\VLD = \SI{210.8}{mV}$.
}\label{fig:triplet-ED_linecuts}
\end{figure}


\begin{figure}[h!]
\centering
\includegraphics[width=\textwidth]{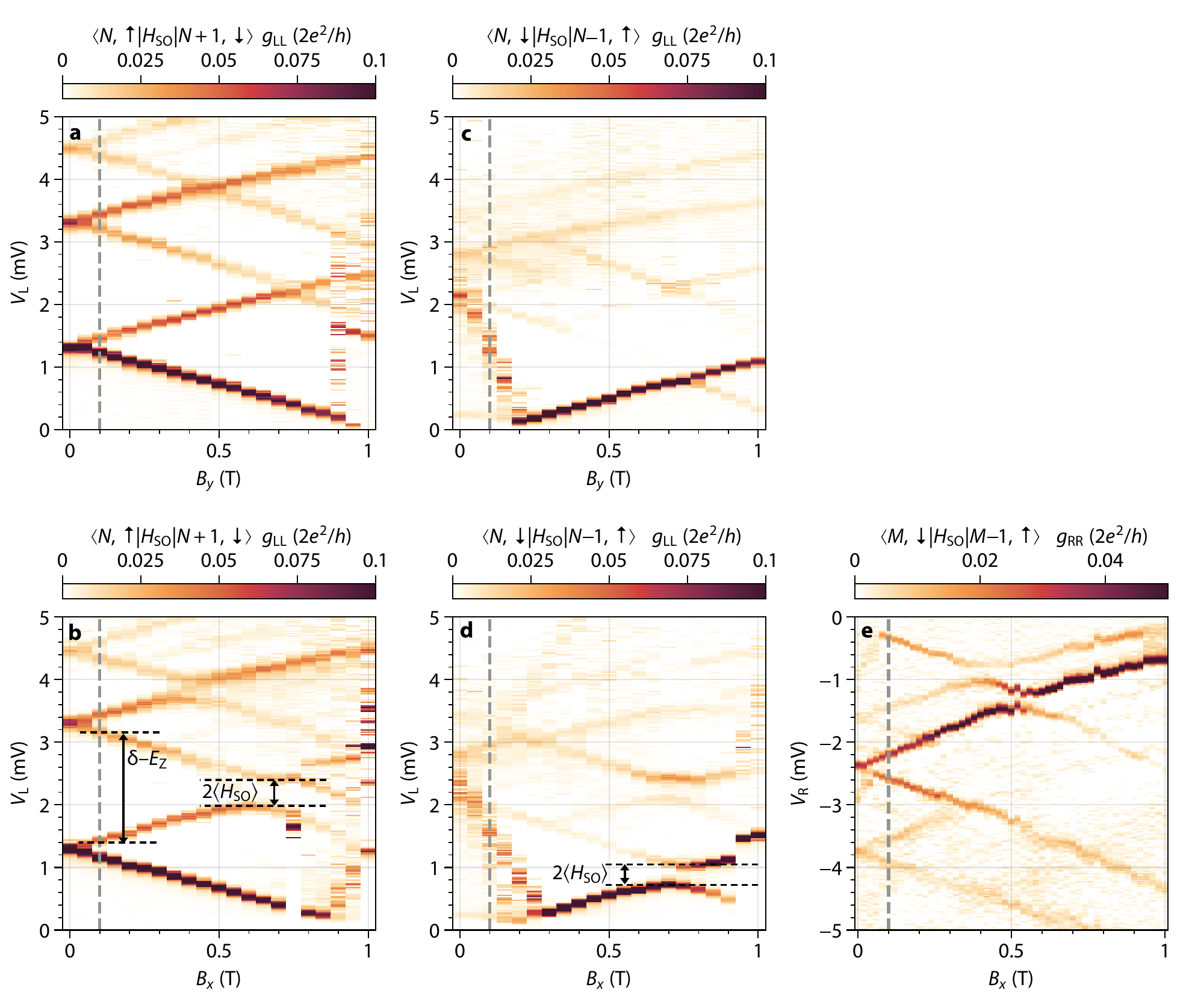}
\caption{\textbf{QD excitation spectra measured using methods described in~\cite{fasth2007direct}, from which we extract QD $g$-factor, level spacing and SOC}.
\textbf{a. b.} Left QD excitation spectra evolving under $B$ applied along $y$ (a) and $x$ (b) for the spin-up ground state. Grey lines mark the field value at which Main Text data are taken. 
The observation that opposite-spin excited states cross each other in panel a means spin is conserved, implying $\BSO$ in the QD and $B$ point along the same direction, i.e., $y$. Opposite-spin states in panel~b, in contrast, anti-cross due to SOC. The quantities needed to calculate the opposite-spin admixture weight (level spacing $\delta$, Zeeman splitting $E_\mathrm{Z}$ and spin-orbit level repulsion gap $2 \langle H_\mathrm{SO}\rangle$ can be directly read from panel~b (see Supplementary Materials for details). Panel~b shows the largest value of spin-orbit level repulsion that we have measured in the QDs, which is used as an upper-bound estimation for the effect of SOC in QD in Supplementary Materials. The Zeeman-splitting slopes yield $g = 45$, i.e., Zeeman energy $g\mu_\mathrm{B} B = \SI{260}{\micro eV}$ at $B = \SI{100}{mT}$.
\textbf{c., d.} Left QD excitation spectra under $B$ along $y$ and $x$ for the spin-down ground state. The $g$-factor and level spacing are similar to those in panels~a,b (as seen in data above \SI{0.3}{T}) but the spin-orbit level repulsion is smaller. 
\textbf{e.} Right QD excitation spectrum under $B$ along $x$ for the spin-up ground state. Anti-crossings of similar widths to panel~d can be observed, although interpretation of the spectrum lines is less clear. No good data could be obtained for the $y$ direction and the spin-down ground state. $\mathrm{d}I/\mathrm{d}V$ in all panels is calculated by taking the numerical derivative after applying a Savitzky-Golay filter of window length 5 and polynomial order 1 to the measured current. The measurements shown here were conducted using different QD orbitals than those used in \Cref{fig:triplet-fig3,fig:triplet-fig4}. The obtained magnitude of the SOC should be taken as an estimate rather than a precise value. 
}\label{fig:triplet-ED_QD_SOI}
\end{figure}


\begin{figure}[h!]
\centering
\includegraphics[width=\textwidth]{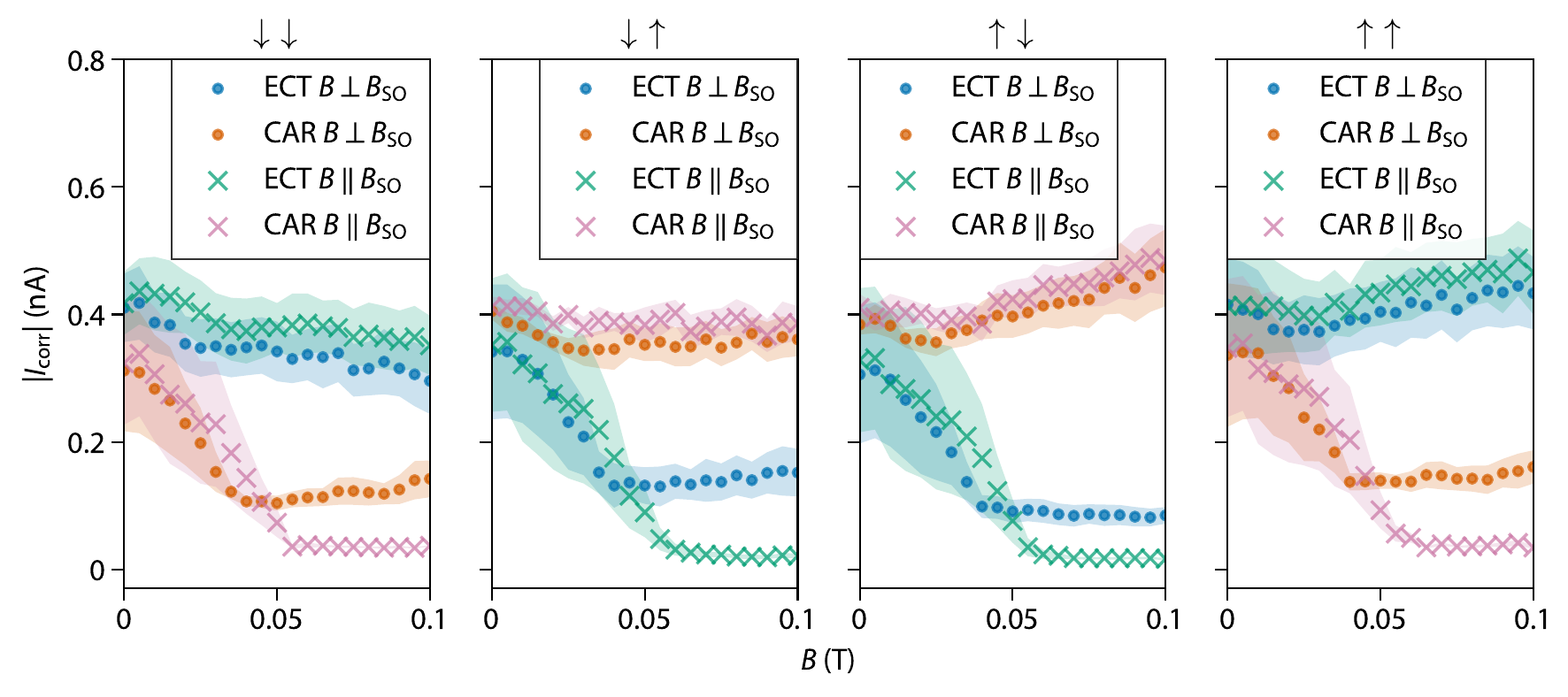}
\caption{\textbf{$B$ dependence of CAR and ECT amplitudes of Device~A.}
Measurements of CAR and ECT at $4\times4$ spin and bias combinations similar to those in \Cref{fig:triplet-fig4}g are performed as functions of $B$, both when $B=B_y \parallel \BSO$ and when $B=B_x \perp \BSO$. At around $\vert B \vert = \SI{50}{mT}$, Zeeman energy exceeds the applied bias voltage of \SI{100}{\micro V} and transport across QDs becomes spin polarized. The equal-spin CAR and opposite-spin ECT amplitudes no longer depend on $\vert B \vert$ significantly at higher fields.
}\label{fig:triplet-ED_B_sweep}
\end{figure}


\begin{figure}[h!]
\centering
\includegraphics[width=0.8\textwidth]{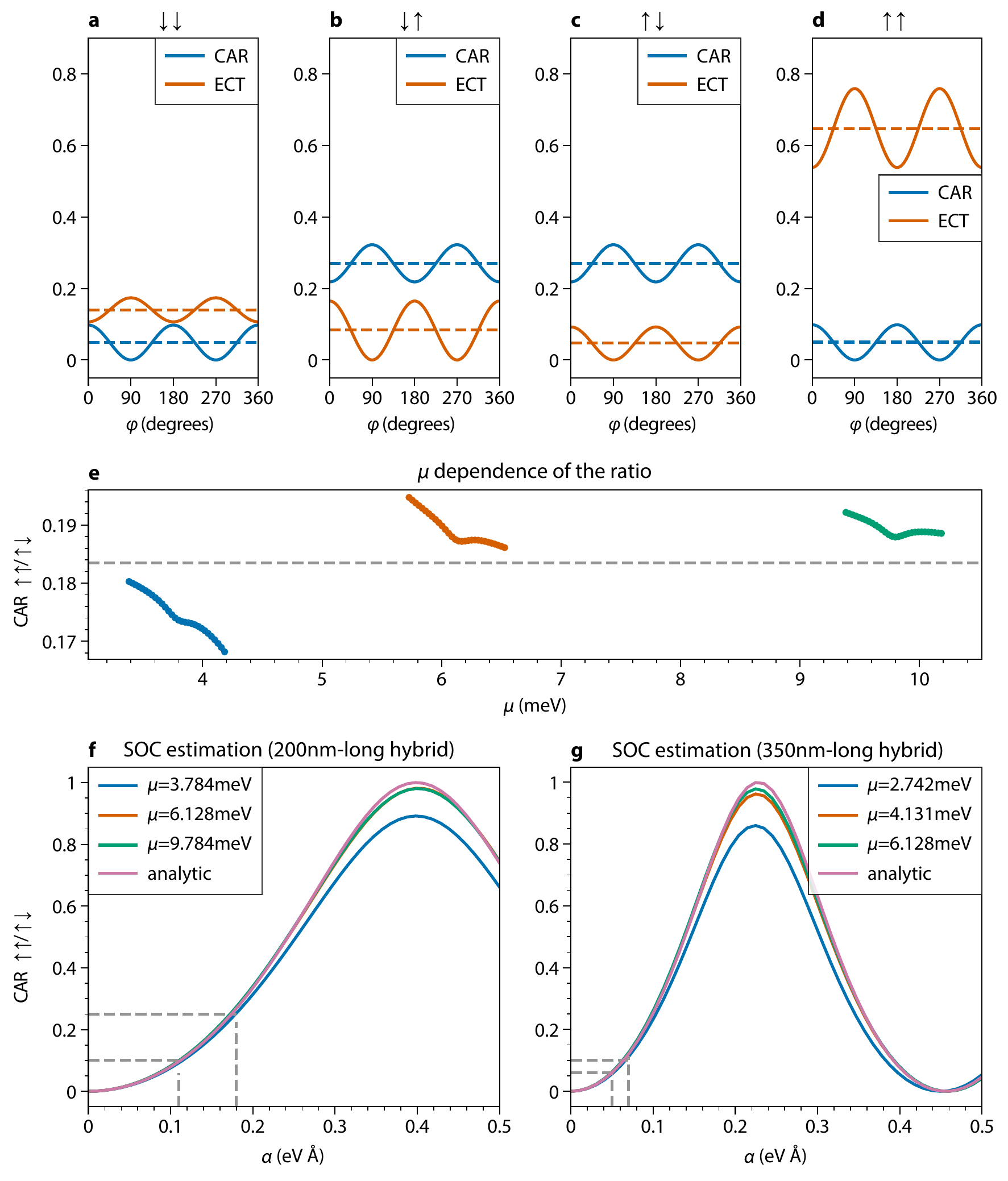}
\caption{\textbf{Theoretical calculations of CAR and ECT amplitudes at finite $B$, from which we extract the SOC strength in the hybrid segment.}
See Supplementary Material and Ref~\cite{liu2022tunable} for details.
\textbf{a.--d.} CAR and ECT amplitudes (proportional to currents) at hybrid-segment $\mu=\SI{6.3}{meV}$ for the four spin combinations when $B$ is rotated in-plane.
Dashed lines are the average of each curve. The ratio between $\uparrow \uparrow$ CAR to $\uparrow \downarrow$ CAR is taken as a proxy of the triplet spin component over singlet in the following panels.
\textbf{e.} Numerical (solid) and analytical (dashed) calculations of angle-averaged $\uparrow\uparrow / \uparrow\downarrow$ CAR ratio are shown in the vicinity of three quantized levels in the hybrid segment (see Supplementary Material and Ref~\cite{liu2022tunable} for details). Variation is small throughout the numerically investigated ranges and all are close to the analytical result, signaling the triplet component estimation is insensitive to the exact chemical potential assumed in the theory. 
\textbf{f.} Dependence of the triplet component on the SOC strength $\alpha$ for a length as in Device~A (\SI{200}{nm}), numerically calculated at three representative chemical potentials together with the analytical result. In \Cref{fig:triplet-fig4}g, triplet/singlet ratios defined here range from $\sim 0.1$ to $\sim 0.25$. This puts the estimation of $\alpha$ in the range of 0.11 to \SI{0.18}{eV \cdot \angstrom}, in agreement with reported values in literature (0.1 to \SI{0.2}{eV \cdot \angstrom} in Refs~\cite{de2018electric, bommer2019spin}). 
\textbf{g.} Dependence of the triplet component on the SOC strength $\alpha$ for a length as in Device~B (\SI{350}{nm}), numerically calculated at three representative chemical potentials together with the analytical result. Similar comparison with data in \Cref{fig:triplet-ED_devB_rot} yields estimations of $\alpha$ in the range of 0.05 to \SI{0.07}{eV \cdot \angstrom}. The weaker SOC could be attributed to the higher $\VPG$ used here (\SI{0.4}{V} for Device~B compared to \SI{0}{V} for Device~A) weakening the inversion-symmetry-breaking electric field.
}\label{fig:triplet-ED_theory}
\end{figure}


\begin{figure}[h!]
\centering
\includegraphics[width=\textwidth]{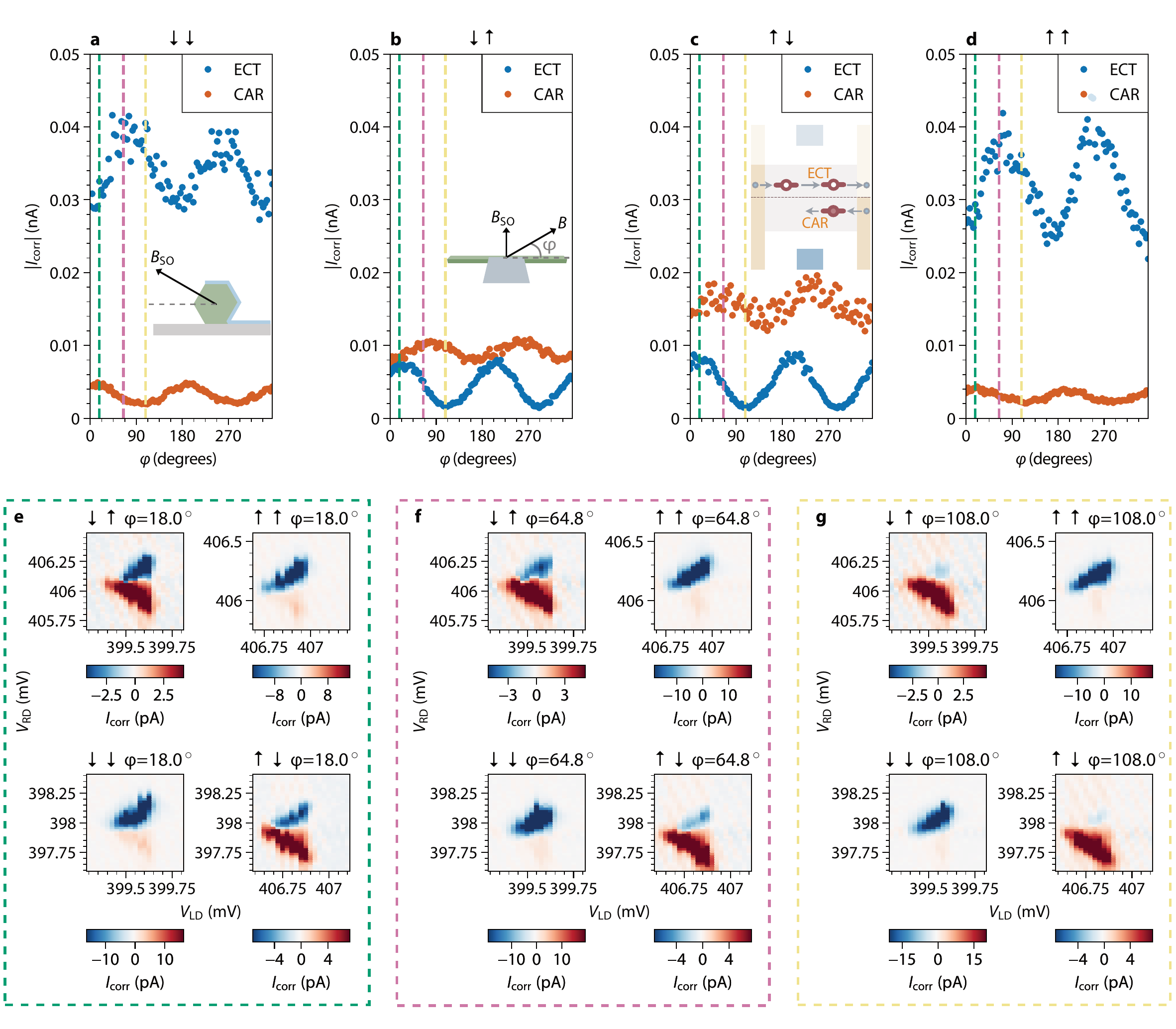}
\caption{\textbf{Anisotropic CAR and ECT reproduced in Device~B}.
Device~B is fabricated similarly except for the absence of the Pt layer to exclude it as a possible spin-flipping mechanism in the nanowire.
\textbf{a.--d.} CAR and ECT amplitudes for four spin combinations when rotating $\vert B \vert =\SI{80}{mT}$ in the plane spanned by the nanowire axis and $\BSO$ (defined as the direction where equal-spin CAR and opposite-spin ECT are maximally suppressed).
The $\BSO$ in this device points approximately $30^\circ$ out of plane (insets: cross-section in a and top view in b).
Inset in panel c: a sketch of the type of bias voltage configurations used in this measurement and in \Cref{fig:triplet-fig4}g; see caption of the lower panels for details.
\textbf{e.--g.} Selected views of $\Icorr$ at three representative angles (marked with boxes of the corresponding color as dashed lines in panels~a--d).
These measurements are performed at $\VL=\SI{70}{\micro V}, \VR=0$ because the right QD allows significant local Andreev current at finite bias due to one malfunctioning gate.
This measurement scheme, which is also employed in Figure~4g, allows us to measure both ECT and CAR without changing the bias.
Inset in panel~c illustrates when CAR and ECT processes occur using $\VL < \VR = 0$ as an example.
Following the same analysis in \Cref{fig:triplet-fig1}, we measure ECT when $-e\VL <\muL=\muR<0$ and CAR when $-e\VL <\muL<0<\muR = -\muL<e\VL$.
The main features of the Main Text data can be reproduced, including anti-diagonal CAR and diagonal ECT lines, strong suppression of opposite-spin ECT and equal-spin CAR along one fixed direction, and their appearance in perpendicular directions.
}\label{fig:triplet-ED_devB_rot}
\end{figure}


\begin{figure}[h!]
\centering
\includegraphics[width=\textwidth]{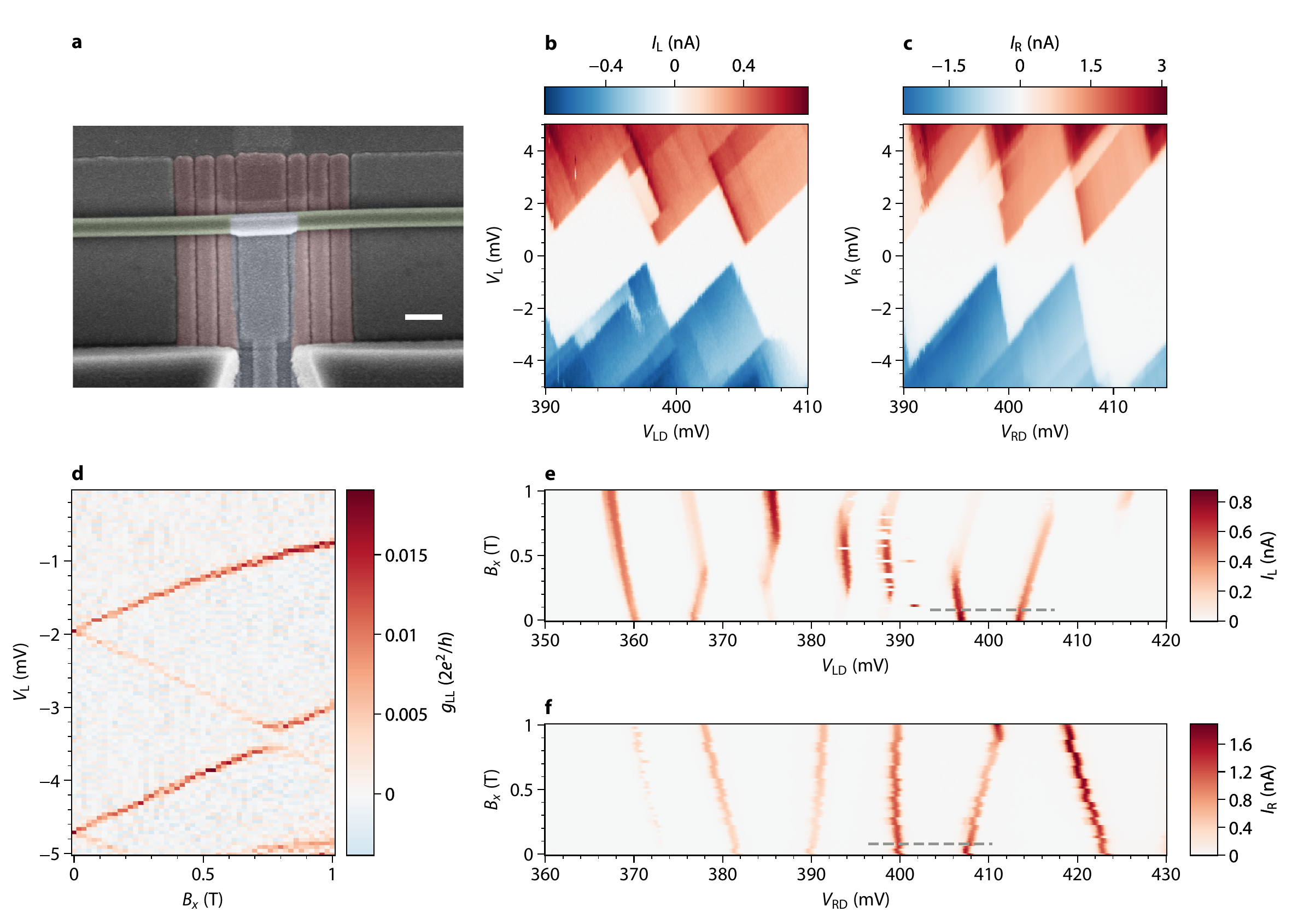}
\caption{\textbf{Device~B characterization.}
\textbf{a.} False-colored SEM image of Device~B prior to the fabrication of N leads. Green is nanowire, blue is Al and red are the bottom gates. Scale bar is \SI{200}{nm}. The hybrid segment is \SI{350}{nm} long.
\textbf{b., c.} QD diamonds of the levels used on both sides at $B=0$.
\textbf{d.} Left QD bias spectroscopy under applied $B=B_x$ and $\VLD = \SI{357}{mV}$ along the nanowire axis.
Level spacing \SI{2.7}{meV}, $g$-factor 61 and spin-orbit anti-crossing $2\langle H_\mathrm{SO} \rangle = \SI{0.25}{meV}$ can be extracted from this plot. $\mathrm{d}I/\mathbf{d}V$ in this panel is calculated by taking the numerical derivative of the measured current.
\textbf{e., f.} Left and right QD levels evolving under finite $B_x$. The levels used for taking the data in \Cref{fig:triplet-ED_devB_rot} and the field at which they are taken are indicated by grey dashed lines. $g$-factor is estimated to be 26 for the right QD.
}\label{fig:triplet-ED_devB_char}
\end{figure}


\begin{figure}[h!]
\centering
\includegraphics[width=0.4\textwidth]{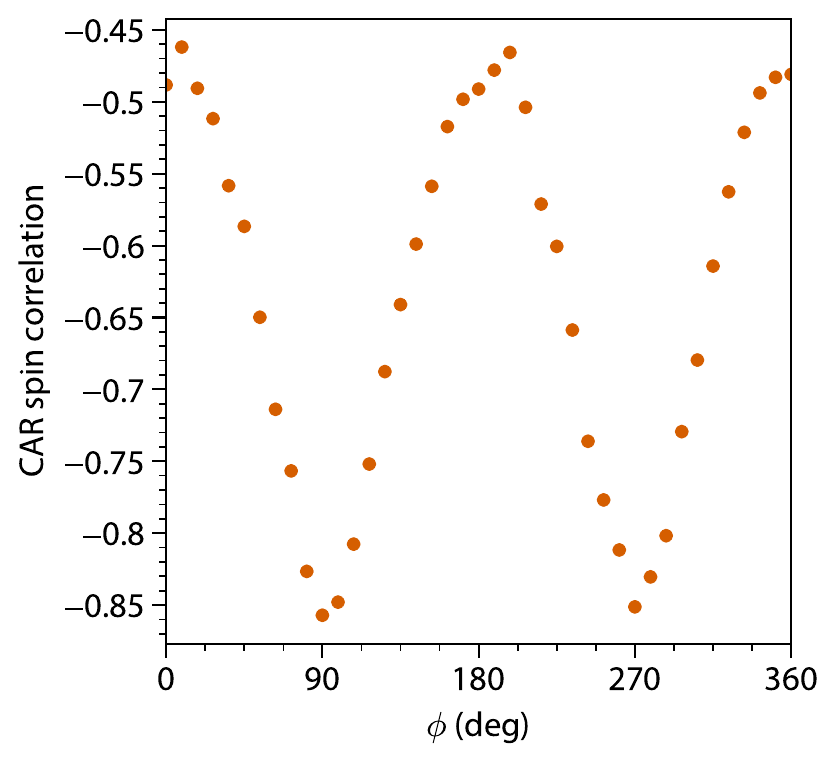}
\caption{\textbf{Spin correlation analysis of the data in \Cref{fig:triplet-fig4}g.}
We define $\mathrm{CAR}^{\uparrow \uparrow}\equiv (\vert \Icorr \vert)$ for the $\uparrow \uparrow$ spin configuration and similarly for the others, as defined in \Cref{fig:triplet-fig4}g. The spin correlation for a given $B$ direction is calculated as $(\mathrm{CAR}^{\uparrow \uparrow} + \mathrm{CAR}^{\downarrow \downarrow} - \mathrm{CAR}^{\uparrow \downarrow} - \mathrm{CAR}^{\downarrow \uparrow}) / (\mathrm{CAR}^{\uparrow \uparrow} + \mathrm{CAR}^{\downarrow \downarrow} + \mathrm{CAR}^{\uparrow \downarrow} + \mathrm{CAR}^{\downarrow \uparrow})$.
Perfectly singlet pairing yields $-1$ spin correlation. The $-0.86$ correlation when $B \parallel \BSO$ is limited by the measurement noise level and can be improved by more signal averaging or more sophisticated analysis methods that are less sensitive to noise.
When $B$ points along other directions, the spin anti-correlation reduces as expected for non-singlet pairing.
}\label{fig:triplet-Spin_corr}
\end{figure}


\begin{figure}[h!]
\centering
\includegraphics[width=0.9\textwidth]{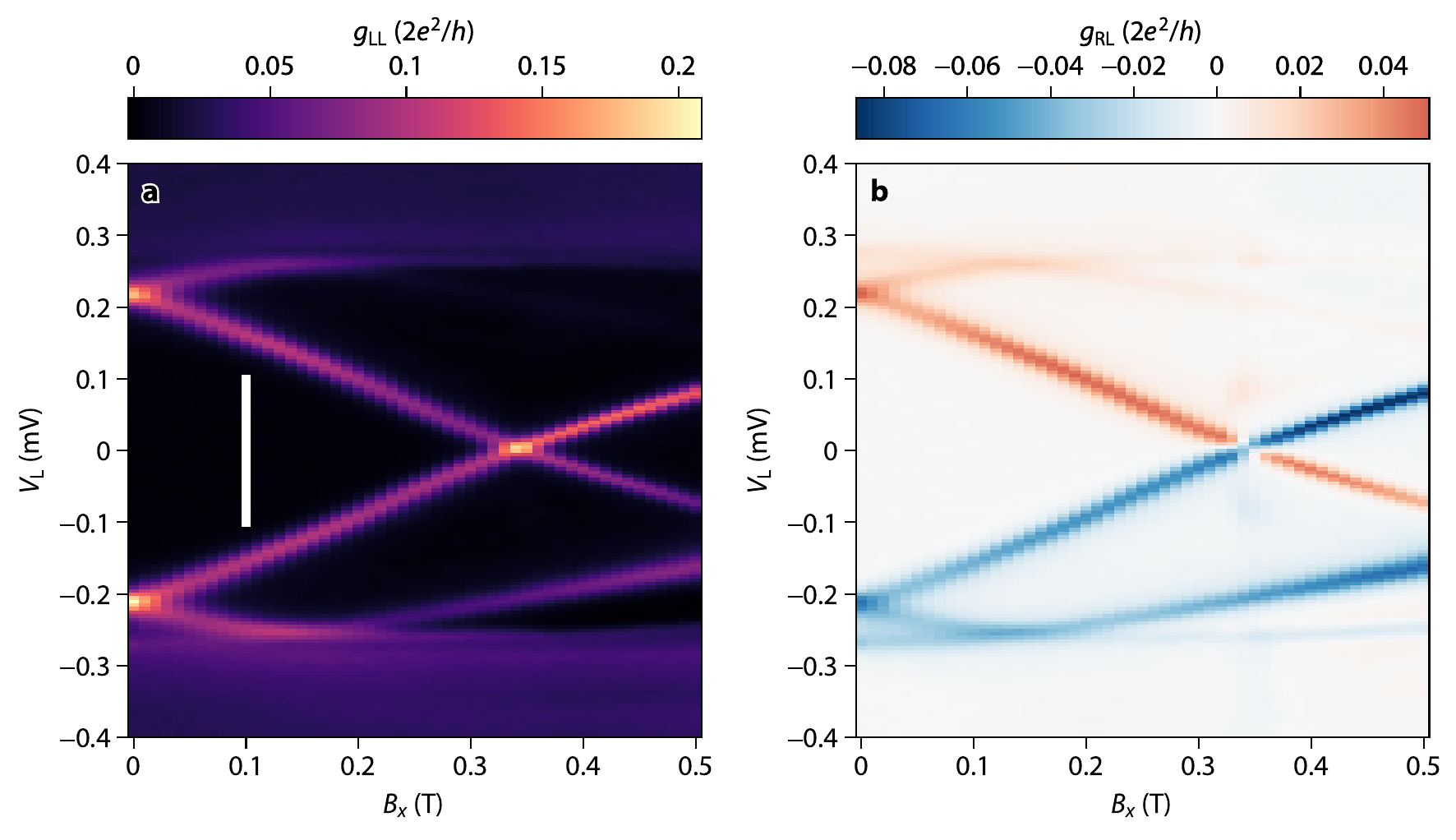}
\caption{\textbf{$B$ dependence of the energy spectrum in the middle hybrid segment of Device~A revealing a discrete Andreev bound state (ABS).}
\textbf{a.} $g_\mathrm{LL} \equiv \mathrm{d}\IL / \mathrm{d} \VL$. White line indicates the bias range in which the experiments at finite $B$ field were performed: the QD energies are kept below the lowest-lying excitation of the middle hybrid segment at all times to avoid sequential tunneling into and out of it.
The $g$-factor of the superconducting-semiconducting hybrid state is seen to be 21 from this plot, smaller than that in QDs.
\textbf{b.} $g_\mathrm{RL}\equiv \mathrm{d}\IR / \mathrm{d} \VL$. The presence of nonlocal conductance corresponding to this state proves this is an extended ABS residing under the entire hybrid segment, tunnel-coupled to both sides.
We note that this is the same dataset presented in another manuscript~\cite{mazur2022spin} where it is argued that the observed Zeeman splitting of this ABS also rules out the possibility of the Pt top layer randomizing spin inside the InSb nanowire.
}\label{fig:triplet-ED_ABS_B}
\end{figure}


\begin{figure}[h!]
\centering
\includegraphics[width=\textwidth]{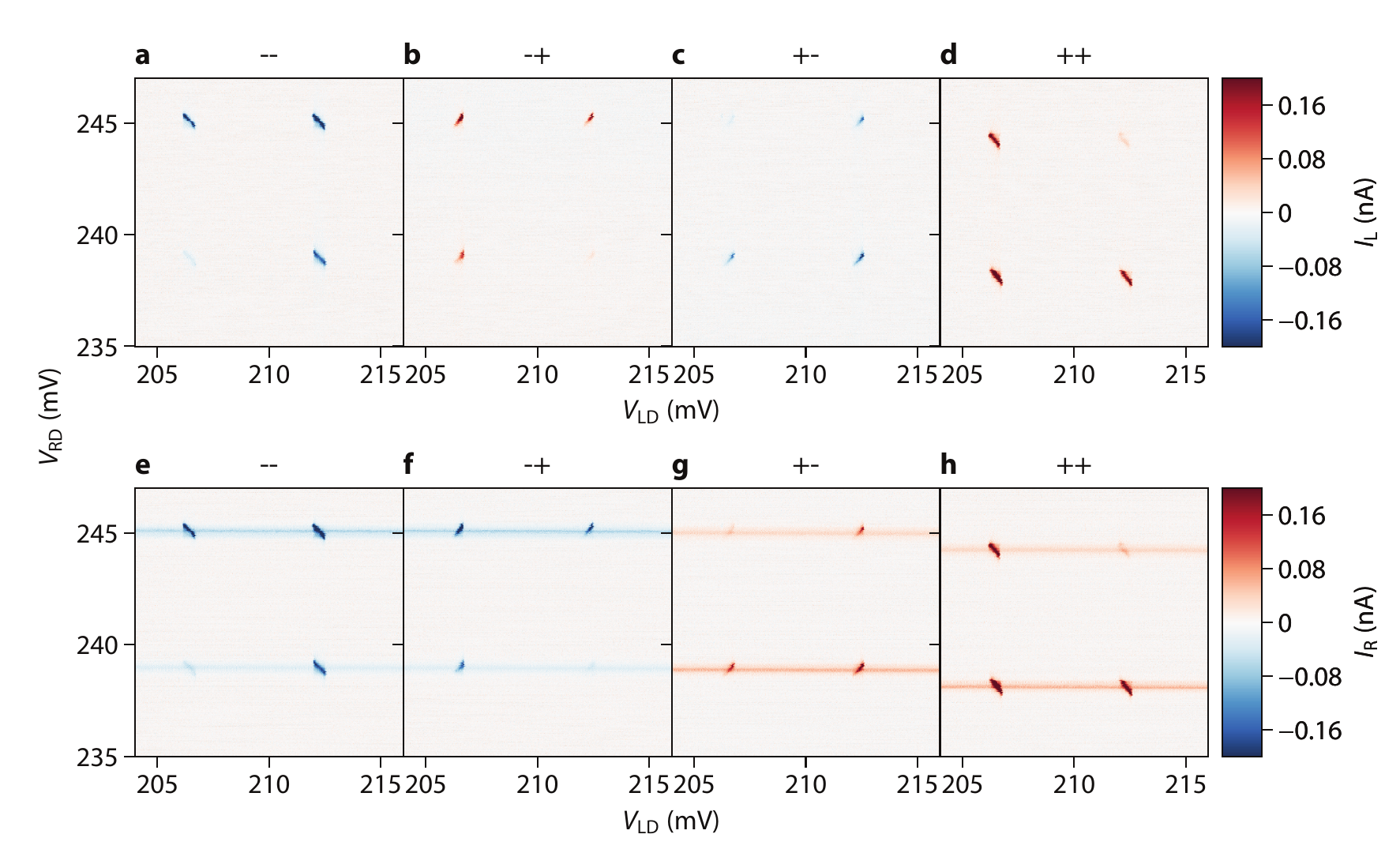}
\caption{\textbf{Plotting of raw data used in \Cref{fig:triplet-fig3}.}
\textbf{a.--h.} $\IL,\IR$ spanning the four joint charge degeneracies and under four N bias polarities at $B =0$. \Cref{fig:triplet-fig3}a, e.g., is obtained by taking data from panels~c and g and calculating their geometric mean at each pixel. The horizontal lines in $\IR$ are due to local Andreev processes carried only by the right junction. Since $\IL=0$ away from the joint charge degeneracies, these purely local currents do not appear in $\Icorr$.
}\label{fig:triplet-ED_raw-1}
\end{figure}

\begin{figure}[h!]
\centering
\includegraphics[width=\textwidth]{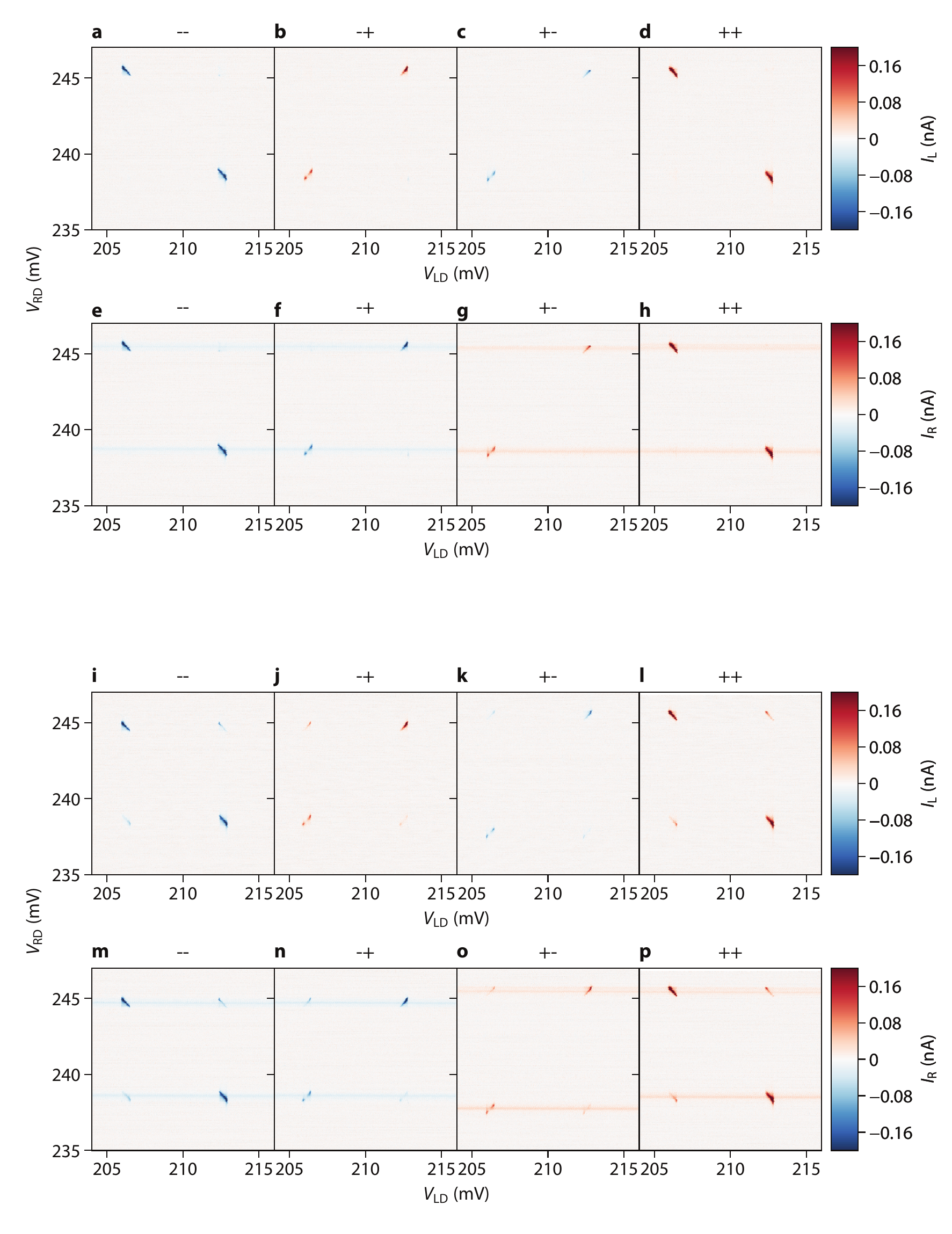}
\caption{\textbf{Plotting of raw data used in \Cref{fig:triplet-fig4}c,~f.}
(For other raw data, see the affiliated data repository.)
\textbf{a.--h.} $\IL,\IR$ spanning the four joint charge degeneracies and under four N bias configurations at $B =B_y=\SI{100}{mT}$.
\textbf{i.--p.} $\IL,\IR$ spanning the four joint charge degeneracies and under four N bias configurations at $B =B_x=\SI{100}{mT}$.
}\label{fig:triplet-ED_raw-2}
\end{figure}

\clearpage

\section{Supplementary material}

\subsection{SOC in QDs}

We have interpreted the measured anisotropy in CAR/ECT amplitudes as resulting from SOC in the hybrid section.
It warrants discussion that the QDs themselves do not have purely spin up or down states in the presence of $B$, either, because of SOC~\cite{hanson2007spins}.
\Cref{fig:triplet-ED_QD_SOI} shows that along $y$, the spin-orbit field direction of the QD, spin is a good quantum number and excited states belonging to different orbitals and opposite spins cross each other. 
When $B=B_x$ is perpendicular to the QD spin-orbit field, QD eigenstates become admixtures of $\ket{n,\uparrow}$ and $\ket{n+1,\downarrow}$ ($n$ being the orbital index number of the QD) and nominally opposite spin states anti-cross.
Qualitatively, this produces the same effect that spin blockade is complete when $B \parallel \BSO$ in both QDs and gets lifted in other directions because of the mixing even if the hybrid segment is purely spin-singlet.
However, the estimation below of the strength of this effect shows it plays a small role in our observed oscillations.

The amount of spin-up mixed into the nominally spin-down QD level is $\langle H_\mathrm{SO} \rangle/(\delta-E_\mathrm{Z})$ according to first-order perturbation theory~\cite{hanson2007spins}, where $2\langle H_\mathrm{SO} \rangle$ is the spin-orbit level repulsion gap between the two mixing orbitals, $\delta$ the orbital level spacing between them and $E_\mathrm{Z}$ the Zeeman splitting. 
All energies can be directly read out from bias spectroscopy of the QD excitation energies as a function of $B$ (\Cref{fig:triplet-ED_QD_SOI}b,d).
The largest value of $\langle H_\mathrm{SO} \rangle/(\delta-E_\mathrm{Z})$ we measured is seen in \Cref{fig:triplet-ED_QD_SOI}b, where the spin-orbit level repulsion is \SI{0.4}{meV} and the difference between Zeeman splitting and level spacing is \SI{1.6}{meV}. 
We thus observe no more than $0.2/1.6=12.5\%$ of the opposite admixture at $B=\SI{100}{mT}$, namely the nominally $\ket{\uparrow}$ state becomes modified as $\ket{\uparrow}+0.125\ket{\downarrow}$ via perturbation.
If for each QD, the nominally spin-up level has a spin composition of $\alpha_\eta \ket{\uparrow} + \beta_\eta \ket{\downarrow}$ and the nominally spin-down is $\beta_\eta \ket{\uparrow} - \alpha_\eta \ket{\downarrow}$ where $\eta=\mathrm{L,R}$ and coefficients are normalized, the spin-flipping probability produced by SOC in QD is $\vert\alpha_\mathrm{L}\beta_\mathrm{R}+\alpha_\mathrm{R}\beta_\mathrm{L}\vert ^2$.
Using the worst-case-scenario numbers extracted from \Cref{fig:triplet-ED_QD_SOI}b above for both QDs, the expected triplet current at $\phi=0^\circ,180^\circ$ should be no more than 6\% of the corresponding singlet value if the superconducting hybrid has no SOC.
Given that our observed triplet current at $\phi=0^\circ,180^\circ$ is between 20\% and 50\% of the spin-singlet counterpart, the majority of the contribution must originate from SOC in the hybrid.
This is corroborated by the fact that a moderate value of SOC taken from the literature for the Al-InSb segment reproduces the measured features and that the maximum spin-blockade direction is roughly $18^\circ$ out of plane for Device~A and $30^\circ$ for Device~B, consistent with the three-facet Al coverage on InSb.
We will also present data on the gate tunability of $\BSO$ in a future manuscript to substantiate this interpretation~\cite{bordin2022controlled}.

The conclusion that SOC effects are much stronger in the hybrid---despite renormalization of InSb parameters by hybridization with Al~\cite{antipov2018effects}---has several possible explanations.
Firstly, the length of the hybrid segment is more than twice the QD size, leading to more spin precession as electrons traverse through it.
In addition, band bending at the Al-InSb interface may lead to much stronger electric fields~\cite{antipov2018effects} than produced by gate electrodes in the QDs.

Although this confirmation of the relative SOC strengths is helpful to asserting the $p$-wave nature of the hybrid superconducting pairing, triplet Cooper pair splitting can indeed also be achieved via two strongly spin-orbit-coupled QDs connected by a purely $s$-wave superconductor under $B$, e.g.~in the situation where the QD dimensions are larger than the hybrid segment.
The key to inducing triplet correlations on two QDs is to introduce non-collinear spin quantization axes in QDs and in the hybrid.
Thus in the semiclassical description, injecting a spin pointing along a different direction than the Zeeman field leads to precession.
The two scenarios are, in fact, physically equivalent in this aspect considering that SOC during tunneling and spatially varying Zeeman quantization axes only differ by a gauge transformation~\cite{braunecker2010spin}.

\subsection{\textit{g}-factor anisotropy}

InSb nanowire QDs have anisotropic $g$-factors~\cite{nadj-perge2012spectroscopy}.
For essentially the same reason as presented in~\cite{nadj-perge2012spectroscopy}, $g$ anisotropy is not a plausible cause of the observed spin blockade and lifting behavior.
In summary, an anisotropic $g$-tensor has three principal axes and rotating $B$ in one plane generally encounters two of them.
Along both axes, the spins in the dots are aligned and blockade is complete. 
This results in two peaks and dips in $180^\circ$ of rotation instead of one, inconsistent with our observed periodicity.

\subsection{Theoretical modelling}

We present here the theoretical model of the experimental data presented in this work.
Details of the model can be found in the parallel theory work~\cite{liu2022tunable}.

\subsubsection{CAR and ECT resonant currents}

In the Main Text, the resonant current flowing through the N-QD-S-QD-N hybrid system is measured.
Such CAR and ECT resonant currents have the forms~\cite{liu2022tunable}
\begin{align} \label{eq:resonant-currents}
&I_{\mathrm{CAR}}= \frac{e}{\hbar} \cdot \frac{\Gamma^2_{DL}}{(\varepsilon_l+\varepsilon_r)^2+\Gamma^2_{DL}} \cdot \frac{\Gamma^2_{\mathrm{CAR}}}{\Gamma_{DL}}, \nn
&I_{\mathrm{ECT}}= \frac{e}{\hbar} \cdot \frac{\Gamma^2_{DL}}{(\varepsilon_l-\varepsilon_r)^2+\Gamma^2_{DL}} \cdot \frac{\Gamma^2_{\mathrm{ECT}}}{\Gamma_{DL}}, 
\end{align}
respectively.
Here $\Gamma_{DL}$ is the total QD-N coupling strength, i.e.\ sum of left-QD--left-N coupling and right-QD--right-N coupling.
$\Gamma_{\mathrm{CAR}}$ and $\Gamma_{\mathrm{ECT}}$ are the effective CAR and ECT coupling between the two QDs. 
These couplings depend on the properties of the middle hybrid segment and on the spin polarization in the QDs, but do not depend on the energies of the QDs.
Thus in the $(\varepsilon_l, \varepsilon_r)$-plane, the resonant current assumes a Lorentzian shape with the broadening being $\Gamma_{DL}$ and reaching the maximum value along $\varepsilon_l= \pm \varepsilon_r$ for ECT and CAR, respectively.
Thereby the maximum current for CAR and ECT are
\begin{align}
&I^{\mathrm{max}}_{a}= \frac{e \Gamma^2_a}{\hbar \Gamma_{DL} }
\propto \Gamma^2_a \propto P_a, 
\end{align}
where $a=\mathrm{CAR}$ or $\mathrm{ECT}$~\cite{liu2022tunable}.
Here $P_{\mathrm{CAR/ECT}}$ are the transition probability of electrons between two dots: $P_{\mathrm{CAR}}$ is the probability of two electrons of a Cooper pair tunneling from the middle superconductor to the two separate dots and $P_{\mathrm{ECT}}$ is the probability of a single electron tunneling from the left to the right QD.

Fitting the correlated currents in \Cref{fig:triplet-fig2} to Equation~\eqref{eq:resonant-currents}, we obtain total QD-N tunnel couplings of 57 (panel~d) and \SI{29}{\micro eV} (panel~e) and QD-QD couplings of 9.6 (panel~d) and \SI{3.8}{\micro eV} (panel~e).

\subsubsection{QD-QD coupling strengths}

The Bogoliubov-de-Gennes (BdG) Hamiltonian for the one-dimensional semiconductor-superconductor hybrid nanowire is
\begin{align}
H_0 = &\half \int^{L}_0 dx \Psi\dg(x) h_0(x) \Psi(x), \nn
h_0(x) =  &\left( -\frac{\hbar^2}{2m^*} \partial_x -\mu \right) \tau_z 
+ i \alpha_R \partial_x  \left( \cos \theta \tau_z \sigma_y +  \sin \theta \sigma_z \right) \nn
&+ E_Z \tau_z \sigma_z
+  \Delta_{\mathrm{ind}} \tau_y \sigma_y,
\label{eq:Ham_0}
\end{align}
where $\Psi(x)=[ c_{\su}(x), c_{\sd}(x), c\dg_{\su}(x), c\dg_{\sd}(x) ]\tp$ is the Nambu spinor, $\tau_{\alpha}$ and $\sigma_{\alpha}$ are Pauli matrices acting on the Nambu and spin space, $L$ is the length of the hybrid nanowire, $m^*$ is the effective mass of InSb, $\mu$ is the chemical potential, $\alpha_R$ is the strength of Rashba spin-orbit coupling, $E_Z $ is half Zeeman spin splitting, $\theta$ is the angle between the spin-orbit field and the Zeeman field, and $\Delta_{\mathrm{ind}}$ is the induced superconducting pairing strength in the nanowire.
In the limit of confinement in all three dimensions as is the case with Device~A, $\Delta_{\mathrm{ind}}$ can be read as the lowest energy that the discrete superconducting-semiconducting hybrid states reach.
We define the direction of the Zeeman field to be always along $\sigma_z$, thus it is the spin-orbit field that rotates in our reference frame.

The tunneling Hamiltonian between the hybrid nanowire and the QDs is
\begin{align}
H_T = \left( -t_l  c\dg_{\eta}(0) d_{l, \eta} -t_r c\dg_{\sigma}(L) d_{r, \sigma} \right) \tau_z + \hc,
\end{align}
where $d_{r, \sigma}$ and $d_{l, \eta}$ are the annihilation operators of the spin-polarized state in QD-$l$ and QD-$r$.
Here we assume that the single-electron tunneling is spin-conserving, and that there is a single QD level with a particular spin polarization (along $\sigma_z$) near the Fermi energy of the normal QDs.
That is, the dot state in the left QD has spin-$\sigma$ and that in the right QD has spin-$\eta$.
In the tunneling limit, i.e., $t_{l,r} \ll \Delta_{\mathrm{ind}}$, the effective CAR or ECT tunneling between the two dots is well described by the second-order virtual process $\langle H_T H^{-1}_0 H_T  \rangle$ where $\langle \cdot \rangle$ denotes the ground state of the hybrid nanowire.
For example, the effective CAR coupling between spin-$\eta$ state in QD-$l$ and spin-$\sigma$ state in QD-$r$ is
\begin{align}
\Gamma_{\mathrm{CAR}, \eta\sigma} &= 
t_l t_r \left(  \langle c\dg_{\eta}(0) {H}^{-1}_0 c\dg_{\sigma}(L)  \rangle -   \langle c\dg_{\sigma}(L)  {H}^{-1}_0 c\dg_{\eta}(0) \rangle \right) \nn
&=\frac{t_l t_r}{\Delta_{\mathrm{ind}}} \sum_{m} \frac{ u_{m \eta}(0) v^*_{m \sigma}(L ) - u_{m \sigma}(L) v^*_{m \eta}(0) }{E_m/\Delta_{\mathrm{ind}}} \nn
&= \frac{t_l t_r}{\Delta_{\mathrm{ind}}} a_{\eta \sigma},
\label{eq:P_CAR}
\end{align}
where $[ u_{m\su}(x), u_{m\sd}(x),v_{m\su}(x),v_{m\sd}(x) ]\tp$ is the Bogoliubov wavefunction of the $m$-th eigenstate with excitation energy $E_m>0$.
Thereby for the resonant current flowing through the normal QDs with particular spin polarizations, we have
\begin{align}
I^{\mathrm{max}}_{\mathrm{CAR}, \eta \sigma} \propto \left( \Gamma_{\mathrm{CAR}, \eta \sigma}\right)^2 \propto  P_{\mathrm{CAR}, \eta \sigma}  = \vert a_{\eta \sigma} \vert^2.
\end{align}
Similarly, for the ECT current, we have
\begin{align}
\Gamma_{\mathrm{ECT}, \eta\sigma} &= 
t_l t_r \left(  \langle c_{\eta}(0) {H}^{-1}_0 c\dg_{\sigma}(L)  \rangle -   \langle  c\dg_{\sigma}(L)   {H}^{-1}_0 c_{\eta}(0) \rangle \right)\nn
&=\frac{t_l t_r}{\Delta_{\mathrm{ind}}}  \sum_{m} \frac{ u_{m \eta}(0)  u^*_{m \sigma}(L ) - v_{m\sigma}(L) v^*_{m \eta}(0) }{E_m/\Delta_{\mathrm{ind}}} \nn
&=\frac{t_l t_r}{\Delta_{\mathrm{ind}}}  b_{\eta\sigma} 
\label{eq:P_ECT}
\end{align}
and 
\begin{align}
I^{\mathrm{max}}_{\mathrm{ECT}, \eta \sigma}\propto \left( \Gamma_{\mathrm{ECT},\eta\sigma}\right)^2 \propto  P_{\mathrm{ECT}, \eta\sigma}   = \vert b_{\eta\sigma}  \vert^2.
\end{align}

For the numerical simulation, we first discretize the Hamiltonian in Eq.~\eqref{eq:Ham_0} into a tight-binding Hamiltonian using KWANT~\cite{Groth2014}.
We then get the eigenenergies and eigenfunctions by diagonalizing the Hamiltonian and calculate the probabilities using Eqs.~\eqref{eq:P_CAR} and \eqref{eq:P_ECT}.

\subsubsection{Parameters in the model}

The physical parameters we choose for the BdG Hamiltonian in Eq.~\eqref{eq:Ham_0} are: 
$L=\SI{200}{nm}$ for Device A and $L=\SI{350}{nm}$ for Device B, $m^*=0.015m_e$ for InSb, $\alpha_R=\SI{0.15}{eV \angstrom}$ as estimated using oscillation amplitudes. 
The bare superconducting gap in the parent superconductor is $\Delta_0=\SI{0.3}{meV}$, extracted from tunnel spectroscopy in \Cref{fig:triplet-ED_ABS_B}.
The renormalized $g$-factor of the hybrid state is estimated to be around 20 from \Cref{fig:triplet-ED_ABS_B}.
This value being half that of InSb also implies the induced superconducting pairing strength $\Delta_\mathrm{ind}$ is around half the parent gap~\cite{antipov2018effects}, around \SI{0.15}{meV}.
At $B=\SI{0.1}{T}$, the estimated Zeeman field strength is $E_Z =\half g \mu_B B\approx \SI{0.06}{meV}$.
After discretizing the continuum Hamiltonian into the lattice model, a chemical potential fluctuation with amplitude $\delta \mu = \SI{10}{meV}$ is added on each site to model a moderate amount of disorder in the nanowire.

\subsubsection{Analytical formula of angular oscillation amplitudes}

The main use of the numerical simulations is to compare with the measured up-up/up-down CAR amplitudes to obtain an estimation of the SOC strength $\alpha_R$.
Ref~\cite{liu2022tunable} also contains derivations of an analytical expression that directly relates $\alpha_R$ to the angle-averaged up-up/up-down ratio of CAR amplitudes:
\begin{equation} \label{eq:soi-analytic}
    \frac{\overline{P}_{\mathrm{CAR},\su \su} }{\overline{P}_{\mathrm{CAR},\su \sd} }
= \frac{\sin^2(k_{so}L)}{2 - \sin^2(k_{so}L) }
\end{equation}
where $k_{so} = m \alpha_R/\hbar^2 $ and $L$ is device length.
This expression is insensitive to microscopic details of the wavefunction in the hybrid produced by varying chemical potential, disorder and inhomogeneous $\alpha_R$. 
It depends only on the nanowire length and the averaged spin-orbit length. 
The ratio is no greater than one and reaches the maximum when the nanowire length is a half-integer multiple of the spin-orbit length.
As we can see in \Cref{fig:triplet-ED_theory}, the difference between numerical simulations and Eq.~\eqref{eq:soi-analytic} is indeed small.
The deviations mostly originate from finite Zeeman field not captured by the analytical formula.
Thus, this simple expression provides us with a way of extracting the spin-orbit constant $\alpha_R$ directly from the anisotropic CAR/ECT measurements without any other knowledge than device length and effective electron mass.

\bibliography{main.bib}

\end{document}